\documentclass[%
reprint,
 amsmath,amssymb,
 aps,
pra,
]{revtex4-1}
\usepackage{amsmath}

\usepackage{graphicx}% Include figure files
\usepackage{subfigure}
\usepackage{dcolumn}% Align table columns on decimal point
\usepackage{bm}
\usepackage{epstopdf}
\usepackage{epsfig}% bold math
\usepackage[colorlinks,
            linkcolor=blue,
            anchorcolor=blue,
            citecolor=blue
            ]{hyperref}
\usepackage{color}

\begin{document}

%\preprint{APS/123-QED}

\title{Exponentially-enhanced Quantum Non-Hermitian Sensing via Optimized Coherent Drive}% Force line breaks with \\
\author{Liying Bao$^{1,2}$, Bo Qi$^{1,2\star}$,  Daoyi Dong$^{3}$%,, Franco Nori$^{4,5}$
\\
$^{1}$\textit{Key Laboratory of Systems and Control, Academy of Mathematics and Systems Science, Chinese Academy of Sciences, Beijing 100190, People's Republic of China}\\
$^{2}$\textit{University of Chinese Academy of Sciences, Beijing 100049, People's Republic of China}\\
$^{3}$\textit{School of Engineering and Information Technology, University of New South Wales, Canberra ACT 2600, Australia}\\
%$^{4}$\textit{Theoretical Quantum Physics Laboratory, RIKEN, Saitama, 351-0198, Japan}\\
%$^{5}$\textit{Physics Department, The University of Michigan, Ann Arbor, Michigan 48109, USA}\\
$^\star$\textit{qibo@amss.ac.cn}}

\date{\today}% It is always \today, today,

\begin{abstract}
Distinct non-Hermitian dynamics has demonstrated its advantages in improving measurement precision over traditional sensing protocols. Multi-mode non-Hermitian lattice dynamics can provide exponentially-enhanced quantum sensing where the quantum Fisher information (QFI) per photon increases exponentially with the lattice size.  However, somewhat surprisingly, it was also shown that the quintessential non-Hermitian skin effect does not provide any true advantage. In this paper, we demonstrate the importance of optimizing the phase of the coherent drive, and the position of the injection and detection in multi-mode non-Hermitian quantum sensing. The QFI per photon can be exponentially-enhanced or exponentially-reduced depending on parameters of the drive and detection. Specifically, it is demonstrated that for large amplification by choosing appropriate coherent drive parameters, the non-Hermitian skin effect \textit{can} provide exponentially-enhanced quantum sensing. Moreover, in the regime beyond linear response,  skin-effect can also provide a dramatic advantage  as compared to the local perturbation, and the proposed protocol is robust in tuning the amplification factor.
%The proposed protocol, which combines ideas from optimal control with the non-Hermitian asymmetry mechanics, is particularly important in practical implementations of non-Hermitian quantum sensing.

\end{abstract}

%\pacs{Valid PACS appear here} % PACS, the Physics and Astronomy
\maketitle

%\tableofcontents

\section{Introduction}
Non-Hermitian systems \cite{ZhangJing2018,Monifi2016,McDonald2020,Budich2020,Koch2021,Okuma2020,Kawabata2020,Park2021,G.-Q.Zhang2021,Howl2021,Roberts2021,Xiao2021,Bao2021,Bensa2021,Rao2021,Zhong2020,Liu2019,Bliokh2019,Pan2020,Cao2020,Scheibner2020,Ashida2020,Sounas2017,Ozdemir2014,Sun2020,Zhangjing2015,Feng2017,Liuyl2017,El-Ganainy2018,Gardiner2000} have been attracting increasing interest, both theoretically and experimentally in diverse fields. One particular concern is whether unconventional properties associated with non-Hermitian systems can be utilized for enhanced quantum sensing. A variety of distinct properties of non-Hermitian systems have been exploited to improve  quantum precision \cite{Demange2012, Jing2014, Chen2016, Liu2016, Chen2017, Langbein2018, Ozdemir2019, Zhang2019, Chen2019, Huai2019, Chu2020, Lau2018, Wiersig2014, Wiersig2016, Hodaei2017, Ren2017, Metelmann2015, Clerk2010}.

The intriguing non-Hermitian degeneracy property known as exceptional point (EP) has been widely investigated \cite{Chen2017, Leykam2017,Zhang2019, Langbein2018, Ozdemir2019, Chen2019, Dembowski2001, Heiss2004, Seyranian2005, Liertzer2012, Heiss2012, Rotter2014, Wiersig2014, Zhen2015, Xu2016, Wiersig2016, Hodaei2017, Sunada2017, Ren2017, Clerk2010}. At EP, not only the eigenenergies but also the eigenstates coalesce. Hence, near the EP the eigenenergies have a diverging susceptibility on small parameter changes. However, to leverage EP, fine tunings of the system parameters are needed. Furthermore, the real effect of EP should be assessed carefully owing to the fact that the coalescence of eigenstates may suppress the diverging susceptibility of eigenenergies \cite{Lau2018,Zhang2019,Chen2019}.

The unconventional non-reciprocity was employed as a powerful resource for quantum sensing  in \cite{Lau2018} when quantum noises are included. It allows to arbitrarily exceed the fundamental limit constraining any reciprocal sensors. In \cite{Bao2021}, by introducing two coherent drives, a uniform bound concerning the best possible measurement rate per photon was set up for both reciprocal and non-reciprocal sensors. It was shown that by  focusing on a two-mode system, the bound is approximately attainable and can, in principle, be made arbitrarily large.

Skin effect is a unique feature of non-Hermitian systems \cite{Okuma2020,Kawabata2020},  which depicts the extreme sensitivity to the boundary conditions. In \cite{Okuma2020}, it was revealed that the skin effect originates from intrinsic non-Hermitian topology, and a unified understanding about the bulk-boundary correspondence and the skin effects in non-Hermitian systems was provided. A novel class of sensors has been introduced by harnessing this striking boundary-sensitivity to make the sensitivity growing exponentially with the size of the device \cite{Budich2020,Koch2021}.  A different exponentially-enhanced protocol was presented in \cite{McDonald2020} by utilizing multi-mode non-Hermitian lattice dynamics and making use of both non-reciprocity and an unusual kind of $\mathbb{Z}_2$  symmetry breaking. With appropriate perturbation Hamiltonian, the protocol yields dramatic enhancement, i.e.,  the quantum Fisher information (QFI) per photon increases exponentially with the lattice size.
However, somewhat surprisingly, in \cite{McDonald2020} it was shown that the quintessential non-Hermitian skin effect (NHSE) does not provide any true advantage in sensing.  To be specific, with the intuitively optimal perturbation Hamiltonian $V_{\text{NHSE}}$,  under which the eigenenergies and eigenstates of the non-Hermitian system are extremely sensitive to changes in boundary conditions implying the skin effect,  the QFI per photon does not show any enhancement as the system size $N$ increases.

Inspired by ideas from control optimization, we demonstrate how to exponentially improve the QFI per photon  by optimizing the excitation signal and the position of the injection (and detection) in multi-mode quantum non-Hermitian sensing.  The importance of optimizing the parameters of the coherent drive and the detection phase is clearly demonstrated by the fact that  the QFI per photon of sensing the same perturbation signal can be exponentially-enhanced or exponentially-reduced relying upon different parameters. To make exponential enhancement of the QFI per photon, the key is to find  proper amplification paths along which the signal is exponentially amplified while the total number of the photons is not. Specifically,  it is shown that  under the same perturbation Hamiltonian $V_{\text{NHSE}}$ as in \cite{McDonald2020}, such amplification paths can be found, which were not included in the analysis in \cite{McDonald2020}. Thus,  different from the conclusion in \cite{McDonald2020}, we demonstrate that the perturbation Hamiltonian $V_{\text{NHSE}}$, which induces skin effect, \textit{can provide exponentially-enhanced sensitivity} if the parameters are judiciously chosen.  Moreover, in the regime beyond linear response, the perturbation $V_{\text{NHSE}}$ can also yield dramatic advantage over the protocol based on a local perturbation as proposed in \cite{McDonald2020}. Furthermore, the enhancement is robust in the sense that the dramatic enhancement can be obtained for a large range of the amplification factor as compared to the case in \cite{McDonald2020}. Thus, our protocol further enriches the research of exponentially-enhanced quantum non-Hermitian sensing  and is particularly important in practical implementations of non-Hermitian quantum sensing.

The paper is organized as follows. In Sec. II, we describe the multi-mode non-Hermitian lattice system and the figure of merit, i.e., signal-to-noise ratio (SNR) per photon. Sec. III focuses on the case where the parameter to be  measured is infinitesimal. We compute the SNR per photon under two different perturbation Hamiltonians. The importance of optimizing parameters for the coherent drive is clearly demonstrated by comparisons and the physical meaning is illustrated by paths of the signal amplification. Sec. IV investigates the case where the parameter to be measured is not infinitesimal so that all the orders of the output field should be considered. Sec. V concludes the paper.

\section{Model and Signal-to-Noise Ratio}
\subsection{A parametric driving non-Hermitian sensor}
\begin{figure}[htbp]
\centering
\subfigure[]{
\includegraphics[scale=0.47]{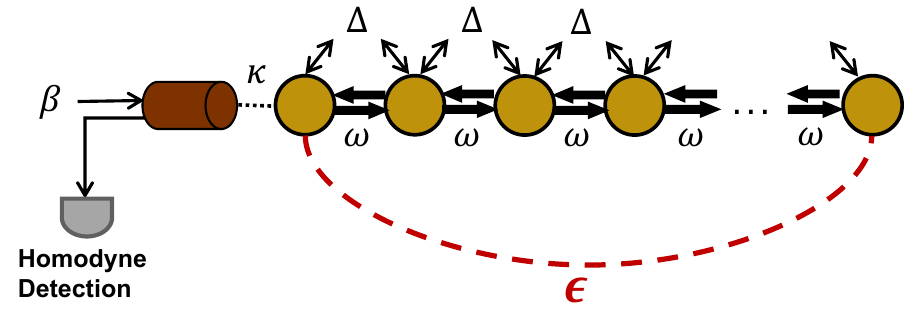} \label{1}
}
\quad
\subfigure[]{
\includegraphics[scale=0.44]{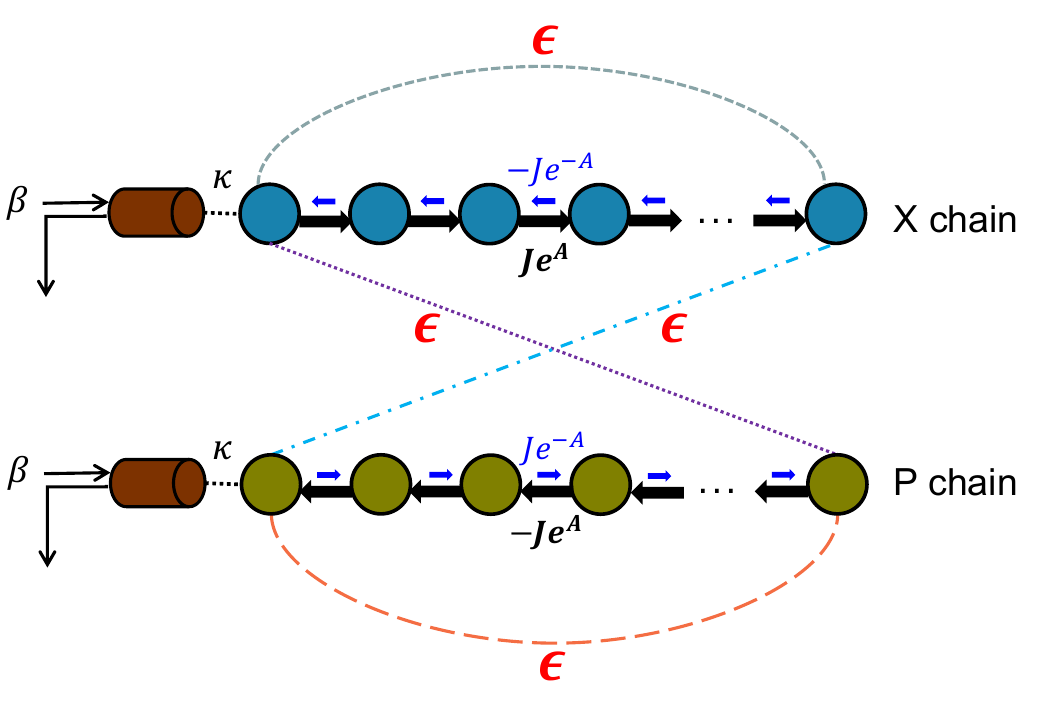} \label{2}
}
\caption{(a) A parametric driving non-Hermitian sensor setup. A 1D chain of $N$ bosonic modes coupled via nearest neighbour hopping $w$ and coherent two-photon drive $\Delta$. There exists a small tunneling between the first mode and the last mode owing to perturbation with amplitude $\epsilon$. A coherent drive $\beta$ is injected into  the chain (at the first site for illustration) through an input-output waveguide with coupling strength $\kappa$.  (b) Two $N$-site non-Hermitian tight-binding Hatano-Nelson chains with effective hopping amplitude $J$ and amplification factor $A$ . The two chains have opposite chiralities for $\hat{x}$ and $\hat{p}$ . For the top (bottom) X (P) chain, hopping to the right is a factor of $e^{2A}$ larger (smaller) than hopping to the left. The first mode and the last mode are coupled due to the presence of small tunneling with amplitude $\epsilon$, allowing the signal to be transmitted between the two chains. There are four possible signal transmission paths in total depicted by different dotted lines with different colors. }
\end{figure}

We adopt the same non-Hermitian sensor setup as in \cite{McDonald2020}.  Consider a $N$-site 1D cavity array subject to parametric drives on each bond described by the Hamiltonian in a rotating frame set by the drive frequency as
\begin{equation}\label{systemhamiltoniana}
\begin{aligned}
\hat{H}_S=\sum^{N-1}_{n=1}(i w \hat{a}^\dagger_{n+1}\hat{a}_n+i \Delta \hat{a}^\dagger_{n+1}\hat{a}^\dagger_n+\textit{h.c.}).
\end{aligned}
\end{equation}
Here we consider open boundary conditions,  and $\hat{a}_i$ denotes the mode annihilation operator on site $i$, $w$ depicts the nearest-neighbour hopping term, $\Delta$ is the nearest-neighbour two-photon drive, and the drive frequency is assumed to be resonant with the cavity mode for simplicity. We take  both $w$ and $\Delta$ to be positive and $w>\Delta$. The Heisenberg equations of motion are
\begin{equation}
\begin{aligned}
\dot{\hat{a}}_n=w \hat{a}_{n-1}+\Delta\hat{a}^\dagger_{n+1}+\Delta\hat{a}^\dagger_{n-1}-w\hat{a}_{n+1}.
\end{aligned}
\end{equation}The setup is illustrated in Fig.~1(a), which may be realized by quantum optical systems or superconducting circuits.

The dynamics generated by $\hat{H}_S$ actually corresponds to two copies of the Hatano-Nelson chain. To make this clear, we turn to the basis of local canonical quadrature operators $\hat{x}_i$ and $\hat{p}_i$ defined by $\hat{a}_i=\frac{\hat{x}_i+i\hat{p}_i}{\sqrt{2}}$, which reads
\begin{equation}
\begin{aligned}
\hat{H}_S=\sum^{N-1}_{n=1}(-(w-\Delta)\hat{x}_{n+1}\hat{p}_{n}+(w+\Delta)\hat{p}_{n+1}\hat{x}_{n}).
\end{aligned}
\end{equation}
The corresponding Heisenberg equations of motion read
\begin{equation}\label{xp}
\begin{aligned}
\dot{\hat{x}}_n&=Je^A\hat{x}_{n-1}-Je^{-A}\hat{x}_{n+1},\\
\dot{\hat{p}}_n&=Je^{-A}\hat{p}_{n-1}-Je^{A}\hat{p}_{n+1},
\end{aligned}
\end{equation}
where the effective hopping amplitude $J$ and amplification factor $A$ (the imaginary vector potential) are defined by
\begin{equation}\label{JA}
J\equiv \sqrt{w^2-\Delta^2},~~~
e^{2A}\equiv \frac{w+\Delta}{w-\Delta}.
\end{equation}
Now it can be seen that each canonical quadrature $\hat{x}$  and $\hat{p}$ corresponds to that of a Hatano-Nelson chain with opposite chirality as illustrated in Fig.~1(b).  As pointed out in \cite{McDonald2020}, when $N$ is even, there is a suppression factor $\frac{\kappa}{2J}$  of susceptibilities as compared with the case where $N$ is odd. Thus, in this paper we only focus on the case with odd  $N$.

To utilize the setup for parameter estimation, we add a perturbation Hamiltonian $\epsilon \hat{V}$ to Hamiltonian $\hat{H}_S$, where $\hat{V}$ is a system operator depicting the perturbation. In this paper,  as in \cite{McDonald2020} we focus on two different kinds of $\hat{V}$. The first one is  $\hat{V}_{\textsf{NHSE}}=e^{i\varphi}\hat{a}^\dagger _1\hat{a}_N+e^{-i\varphi}\hat{a}_1\hat{a}^\dagger_N$, which induces a small tunneling with strength $\epsilon$ between the first and last sites. With this perturbation, the eigenenergies and eigenstates of the non-Hermitian system are extremely sensitive to changes in boundary conditions implying the skin effect. The other one is $\hat{V}_N=\hat{a}_N^{\dag}\hat{a}_N$  representing a local perturbation at site $N$, and $\epsilon$ describes a small change in the resonance frequency of the last site.

In addition, a coherent excitation signal $\beta$ is injected into the chain through
an input-output waveguide coupled to the  $m$-th mode of the chain (labelled in order from the leftmost). In Fig.~1(a), the coherent drive is injected into the first site as an illustration. The inflected signal is measured by a Homodyne detection.  The effective system Hamiltonian reads
\begin{equation}\label{fullHamiltonian}
\begin{aligned}
\hat{H}[\epsilon]=\hat{H}_S+\epsilon\hat{V}+\hat{H}_\kappa-i\sqrt{\kappa}(\hat{a}_m^\dagger \beta-\hat{a}_{m}\beta^\dagger),
\end{aligned}
\end{equation}
where $\hat{H}_\kappa$ describes the damping of the site $m$ at a rate $\kappa$ owing to the coupling with the waveguide \cite{Clerk2010,Gardiner2000}, whose details can be found in Appendix A.  The noise $\hat{B}^{\textsf{in}}(t)$ denotes the accompanied quantum noise of the coherent drive  entering through the waveguide. To ensure the Markovian nature of the full dynamics, $\hat{B}^{\textsf{in}}(t)$ is assumed to be quantum Gaussian \cite{Gardiner2000} satisfying
\begin{equation}
\langle \hat{B}^{\textsf{in}}(t)\hat{B}^{\textsf{in}\dagger}(t')\rangle=(\bar{n}_{\textsf{th}}+1)\delta(t-t'), \end{equation}
\begin{equation}
\langle \hat{B}^{\textsf{in}\dagger}(t)\hat{B}^{\textsf{in}}(t')\rangle=\bar{n}_{\textsf{th}}\delta(t-t'), \end{equation}
\begin{equation}
\langle \hat{B}^{\textsf{in}}(t)\hat{B}^{\textsf{in}}(t')\rangle=0,
\end{equation}
where $\bar{n}_{\textsf{th}}$ is the number of thermal quanta in the input field and $\langle\cdot\rangle$ represents the mean over the state of the bath degrees of freedom. We focus on the case where $\bar{n}_{\textsf{th}}=0$ and it can be directly generalized to classical cases where $\bar{n}_{\textsf{th}}\gg 1$.

\subsection{Signal-to-noise ratio per photon }

From the standard input-output theory \cite{Clerk2010}, the output field is described as
\begin{equation}\label{Bout}
\begin{aligned}
\hat{B}^{\textsf{out}}(t)=\beta+\hat{B}^{\textsf{in}}(t)+\sqrt{\kappa}\hat{a}_m(t).
\end{aligned}
\end{equation}
Our aim is to estimate the small perturbation $\epsilon$ by performing an optimal measurement on the output field $\hat{B}^{\textsf{out}}(t)$.

We first focus on the case where $\epsilon$ is infinitesimal. Note that our system is stable as long as $\kappa>0$. To further ensure the steady state being unique, the site $m$  through which  the coherent drive is injected should be odd. Then in order to estimate an infinitesimal $\epsilon$, we can integrate the output field over a long time period $[0, \tau]$. The temporal mode of the output field can be described as
\begin{equation}\label{Bout(N)}
\begin{aligned}
\hat{\mathcal{B}}_\tau(N)=\frac{1}{\sqrt{\tau}}\int^\tau_0dt \hat{B}^{\textsf{out}}(t),
\end{aligned}
\end{equation}
satisfying $[\hat{\mathcal{B}}_\tau(N),\hat{\mathcal{B}}^\dagger_\tau(N)]=1$. Thus, $\hat{\mathcal{B}}_\tau(N)$ is a canonical bosonic annihilation operator. Since we are interested in how the measurement precision depends on the chain scale $N$, the chain size $N$ is explicitly written here.

The precision of parameter estimation is generally expressed by the standard deviation
\begin{equation}
\delta\epsilon\equiv\sqrt{\langle (\epsilon_{\text{est}}-\epsilon)^2\rangle}.
\end{equation}
It is well-known that, for unbiased estimators, the quantum Cram$\acute{\textrm{e}}$r-Rao inequality sets a lower bound on the precision in terms of the QFI as
\begin{equation}
\delta\epsilon\geq \frac{1}{\sqrt{\nu ~\text{QFI}}},
\end{equation}
where $\nu$ is the round of measurements. To attain the bound, an optimal measurement has to be performed. In general, it is not easy to identify and implement the optimal measurement. However, in the linear Gaussian systems as in Sec. II A, if the amplitude of the coherent drive $|\beta|$ is sufficiently large, it can be verified that
the optimal measurement always corresponds to a standard Homodyne measurement \cite{Banchi2015,Pinel2013}, in the form of
\begin{equation}\label{M(N)}
\begin{aligned}
\hat{\mathcal{M}}_\tau(N)=\frac{1}{\sqrt{2}}\Big{(}e^{-i\phi}\hat{\mathcal{B}}_\tau(N)+e^{i\phi}\hat{\mathcal{B}}^\dagger_\tau(N)\Big{)},
\end{aligned}
\end{equation}
where $\phi\in[0,~\frac{\pi}{2}]$ is called the measurement angle.

Define the signal power and noise power in terms of $\hat{\mathcal{M}}_\tau(N)$ as
\begin{equation}\label{signal}
\begin{aligned}
\mathcal{S}_\tau(N,\epsilon)&=|\langle\hat{\mathcal{M}}_\tau(N)\rangle_\epsilon-\langle\hat{\mathcal{M}}_\tau(N)\rangle_0|^2,
\end{aligned}
\end{equation}
and
\begin{equation}\label{noise}
\begin{aligned}
\mathcal{N}_\tau(N,\epsilon)&=\langle\hat{\mathcal{M}}_\tau^2(N)\rangle_\epsilon-\langle\hat{\mathcal{M}}_\tau(N)\rangle_\epsilon^2,
\end{aligned}
\end{equation}
respectively. The average $\langle\cdot\rangle_\epsilon$ represents the mean with respect to a steady state whose dynamics is governed by $\hat{H}[\epsilon]$. Moreover, define the signal-to-noise ratio as
\begin{equation}
\textrm{SNR}_{\tau}(N)\equiv\frac{\mathcal{S}_\tau(N,\epsilon)}{\mathcal{N}_\tau(N,\epsilon)}.
\end{equation}

It can be verified (see Appendix B) that in the large drive limit ($|\beta|\gg1$), QFI has the following relationship with $\textrm{SNR}_{\tau}(N)$ \cite{Banchi2015,Lau2018}{\color{blue}} as
\begin{equation}\label{QFI}
\begin{aligned}
\underset{\epsilon\rightarrow 0}{\lim}\textrm{QFI}_\tau(N)=\underset{\phi}{\max}[\underset{\epsilon\rightarrow 0}{\lim}
\frac{\textrm{SNR}_{\tau}(N)}{\epsilon^2}].
\end{aligned}
\end{equation}
This implies that the dominate terms of the SNR and  $\epsilon^2\text{QFI}$  are the same for an infinitesimal perturbation $\epsilon \hat{V}$, and thus we can adopt SNR instead of QFI as a figure of merit in the following. However, to make comparisons between different sensors fair, we need to further constrain  resources used during the measurement. In this paper, as in \cite{McDonald2020,Lau2018,Bao2021} we take the SNR per photon
\begin{equation}
\overline{\textrm{SNR}}_{\tau}(N)\equiv\frac{\textrm{SNR}_\tau(N)}{\bar{n}_{\textsf{tot}}(0)}
\end{equation}
as the figure of merit, where the total average photon number
\begin{equation}\label{totalphoton}
\begin{aligned}
\bar{n}_{\textsf{tot}}(0)\equiv \sum_{n} \langle\hat{a}_n^\dagger\hat{a}_n\rangle_0\simeq \sum_n \langle\hat{a}_n^\dagger\rangle_0\langle\hat{a}_n\rangle_0.
\end{aligned}
\end{equation}
Here since $\epsilon$ is infinitesimal, we only consider the zeroth order of $\bar{n}_{\textsf{tot}}$ with respect to $\epsilon$.
Note that  Eq.~(\ref{totalphoton}) is valid in the large $|\beta|$ limit. This is reasonable because the incoherent photons injected by the bath are independent of the coherent drive $\beta$. If the coherent drive is sufficiently large, the coherent drive-induced photons dominate the total average photon number.

From Eqs.~\eqref{Bout(N)}, \eqref{M(N)} and~\eqref{signal}, the first order of the signal in $\epsilon$  is
\begin{equation}\label{signala}
\begin{aligned}
\mathcal{S}_\tau(N,\epsilon)=2\kappa\tau\Big{|}\textrm{Re}[e^{-i\phi}\delta\langle\hat{a}_m\rangle]\Big{|}^2,
\end{aligned}
\end{equation}
where
\begin{equation}
\begin{aligned}
\delta\langle\hat{a}_m\rangle\equiv\epsilon \mathop{\textrm{lim}}\limits_{\epsilon\rightarrow0} \frac{\langle\hat{a}_m\rangle_\epsilon-\langle\hat{a}_m\rangle_0}{\epsilon}
\end{aligned}
\end{equation}
depends on the specific form of the perturbation Hamiltonian $\hat{V}$. Moreover,  it is clear that $\textrm{QFI}_\tau(N)$  is dependent on the first order of $\mathcal{S}_\tau(N,\epsilon)$ with respect to $\epsilon$ and the zeroth order of $\mathcal{N}_\tau(N,\epsilon)$ in $\epsilon$.

Noting that the expression of $\overline{\textrm{SNR}}_{\tau}(N)$ still depends on the input phase $\theta$ of the coherent drive $\beta=|\beta|e^{i\theta}$, the input position $m$, the form of the perturbation Hamiltonian $\hat{V}$, as well as the measurement angle $\phi$. In the following, we will optimize them to obtain a good sensor that can greatly improve the sensitivity.

\section{Exponential enhancement of $\overline{\textrm{SNR}}_{\tau}(N)$  via parameters optimization}

We now calculate  $\overline{\textrm{SNR}}_{\tau}(N)$ in the case of $\epsilon\rightarrow 0$ limit. To do this, we first focus
on $\hat{V}_{\textsf{NHSE}}=e^{i\varphi}\hat{a}^\dagger _1\hat{a}_N+e^{-i\varphi}\hat{a}_1\hat{a}^\dagger_N$ which is related to the NHSE, and then investigate  $\hat{V}_N=\hat{a}_N^{\dag}\hat{a}_N$ which  represents a local perturbation at site $N$.

\subsection{$\overline{\textrm{SNR}}_{\tau}(N)$ with $\hat{V}_{\textsf{NHSE}}$ }

From Eq.~(\ref{xp}), $\hat{H}_S$ consists of two copies of the Hatano-Nelson chain, which have a strong sensitivity to changes in boundary conditions. Thus intuitively, the perturbation $\hat{V}_{\textsf{NHSE}}$ which induces tunneling between the first and last sites would be the optimal perturbation Hamiltonian to sense an unknown parameter $\epsilon$.  In fact, with $\hat{V}_{\textsf{NHSE}}$, the eigenvalues of the system matrix have a strong sensitivity to the perturbation strength $\epsilon$, i.e.,  the non-Hermitian system exhibits skin effect. However, by utilizing the setup in \cite{McDonald2020}, where a real drive ($\theta=0$) is injected into the chain at site 1 ($m=1$), it was shown that  $\hat{V}_{\textsf{NHSE}}$ does not provide a true advantage in sensing. In this subsection, we find that some paths of amplifying the signal  were not included in \cite{McDonald2020}, and then demonstrate the importance of choosing appropriate parameters for the coherent drive in quantum non-Hermitian sensing by computing $\overline{\textrm{SNR}}_{\tau}(N)$ with different parameters of the drive.

According to the non-Hermitian Hamiltonian Eq.~\eqref{fullHamiltonian} and $\hat{V}_{\textsf{NHSE}}=e^{i\varphi}\hat{a}^\dagger _1\hat{a}_N+e^{-i\varphi}\hat{a}_1\hat{a}^\dagger_N$, the full dynamics can be described by the Heisenberg-Langevin equations in terms of $\hat{x}$ and $\hat{p}$ quadratures as (see Appendix C for detailed derivation)
\begin{equation}\label{xpv1}
\begin{aligned}
\dot{\hat{x}}_n=&-Je^{-A}\hat{x}_{n+1}+Je^{A}\hat{x}_{n-1}-\frac{\kappa}{2}\hat{x}_m\delta_{nm}\\
&+\epsilon\textrm{cos}\varphi(\hat{p}_N\delta_{n1}+\hat{p}_1\delta_{nN})-\epsilon \textrm{sin}\varphi(\hat{x}_1\delta_{nN}-\hat{x}_N\delta_{n1})\\ &-\sqrt{2\kappa}|\beta|\textrm{cos}\theta\delta_{nm}-\sqrt{\kappa}\hat{X}^{\textrm{in}}\delta_{nm},\\
\dot{\hat{p}}_n=&Je^{-A}\hat{p}_{n-1}-Je^A\hat{p}_{n+1}-\frac{\kappa}{2}\hat{p}_m\delta_{nm}\\
&-\epsilon\textrm{cos}\varphi(\hat{x}_N\delta_{n1}+\hat{x}_1\delta_{nN})-\epsilon\textrm{sin}\varphi(\hat{p}_1\delta_{nN}-\hat{p}_N\delta_{n1})\\
&-\sqrt{2\kappa}|\beta|\textrm{sin}\theta\delta_{nm}-\sqrt{\kappa}\hat{P}^{\textrm{in}}\delta_{nm},
\end{aligned}
\end{equation}
where $\hat{X}^{\textsf{in}}$ and $\hat{P}^{\textsf{in}}$ are defined via $\hat{B}^{\textsf{in}}=\frac{\hat{X}^{\textsf{in}}+i\hat{P}^{\textsf{in}}}{\sqrt{2}}$. The terms containing $\epsilon$ in Eq.~(\ref{xpv1}) depict the coupling between the quadratures of site 1 and site $N$ owing to the perturbation $\hat{V}_{\textsf{NHSE}}$. For example, a wavepacket can \textit{first} tunnel between the first and last sites with possible quadrature change, and then propagate along the corresponding chain. In \cite{McDonald2020}, such possible paths were not considered, and they only discussed the case where a wavepacket first travels along the chain and then tunnels between the first and last sites. However, it is demonstrated in the following that these paths are crucial in obtaining exponentially-enhanced SNR per photon.

The average values of  $\hat{X}^{\textsf{in}}$ and $\hat{P}^{\textsf{in}}$ are zero, and their second moments satisfy
\begin{equation}
\langle \hat{X}^{\textsf{in}}(t)\hat{X}^{\textsf{in}}(t')\rangle=(\bar{n}_{\textsf{th}}+\frac{1}{2})\delta(t-t'), \end{equation}
\begin{equation}
\langle \hat{P}^{\textsf{in}}(t)\hat{P}^{\textsf{in}}(t')\rangle=(\bar{n}_{\textsf{th}}+\frac{1}{2})\delta(t-t'), \end{equation}
\begin{equation}
\frac{1}{2}\langle \{ \hat{X}^{\textsf{in}}(t),\hat{P}^{\textsf{in}}(t')\}\rangle=0.
\end{equation}

Details of calculating $\overline{\textrm{SNR}}_{\tau}(N)$  on the basis of Eqs.~(\ref{M(N)}) and (\ref{xpv1}) can be found in Appendix C.   We compute the dominate term of $\overline{\textrm{SNR}}_{\tau}(N)$ as
\begin{equation}
\overline{\textrm{SNR}}_\textsf{D}(N)\equiv\frac{\mathcal{S}_\tau(N,\epsilon)}
{\mathcal{N}_\tau(N,0)}\cdot\frac{1}{\bar{n}_{\textsf{tot,D}}(0)},
\end{equation}
where $\bar{n}_{\textsf{tot,D}}(0)$
is the dominate term of $\bar{n}_{\textsf{tot}}(0)$. This consideration of approximation is reasonable because for large amplification factor $A$, it satisfies (see Appendix C)
\begin{equation}
\overline{\textrm{SNR}}_{\tau}(N)=\overline{\textrm{SNR}}_\textsf{D}(N)\cdot\Big{(}1+O(e^{-2A})\Big{)}.
\end{equation}
Now let us investigate $\overline{\textrm{SNR}}_\textsf{D}$  with different coherent drive parameters.

%\begin{figure*}[htbp]
%\subfigure[$~\theta=0,~m=1,~\phi\in {[{0,\frac{\pi}{2}}]}$:
%$\mathcal{S}_\tau(N,\epsilon)\propto e^{2A{(N-1)}},~\bar{n}_{\textsf{tot}}\propto e^{2A{(N-1)}}$, ~~~~~~~~~~~~$\overline{\textrm{SNR}}_\textsf{D}(N)\propto O(1)$.~~~~~]{
%\includegraphics[scale=0.5]{2_1} \label{3}
%}
%\quad
%\subfigure[$~\theta=0,~m=N,~\phi \in {[{0,\frac{\pi}{2}})}$: $\mathcal{S}_\tau(N,\epsilon)\propto e^{2A{(N-1)}},~\bar{n}_{\textsf{tot}}\propto O(1)$, $\overline{\textrm{SNR}}_\textsf{D}(N)\propto e^{2A(N-1)}$.]{
%\includegraphics[scale=0.5]{2_2} \label{4}
%}
%\quad
%\subfigure[$~\theta=\frac{\pi}{2},~m=1~,~\phi \in {({0,\frac{\pi}{2}}]}$:
%~~~~~~~~~$\mathcal{S}_\tau(N,\epsilon)\propto e^{2A{(N-1)}},~\bar{n}_{\textsf{tot}}\propto O(1)$, ~~~~~~~~~~~~~~~~~~~$\overline{\textrm{SNR}}_\textsf{D}(N)\propto e^{2A(N-1)}$.]{
%\includegraphics[scale=0.5]{2_3} \label{5}
%}
%\quad
%\subfigure[$~\theta=\frac{\pi}{2},~m=N,~\phi \in {[{0,\frac{\pi}{2}}]}$: ~~~$\mathcal{S}_\tau(N,\epsilon)\propto e^{2A{(N-1)}},~\bar{n}_{\textsf{tot}}\propto e^{2A{(N-1)}}$, ~$\overline{\textrm{SNR}}_\textsf{D}(N)\propto O(1).$]{
%\includegraphics[scale=0.5]{2_4} \label{6}
%}
%\caption{$\overline{\textrm{SNR}}_\textsf{D}$ with different values of $\theta$, $m$ and $\phi$ for $\hat{V}_{\textsf{NHSE}}$. The paths amplifying the signal power are illustrated in each case.}
%\end{figure*}

\begin{figure*}[htbp]
\subfigure[$~\theta=0,~m=1,~\phi\in {[{0,\frac{\pi}{2}}]}:~~~~~~~~~~~~~~~$
$~~~~~~~~\mathcal{S}_\tau(N,\epsilon)\propto e^{2A{(N-1)}},~\bar{n}_{\textsf{tot}}\propto e^{2A{(N-1)}}$, ~~~~~~~~\leftline{~~~~~~~~~~~~~~~~~~~~~$\overline{\textrm{SNR}}_\textsf{D}(N)\propto O(1)$.}]{
\includegraphics[scale=0.48]{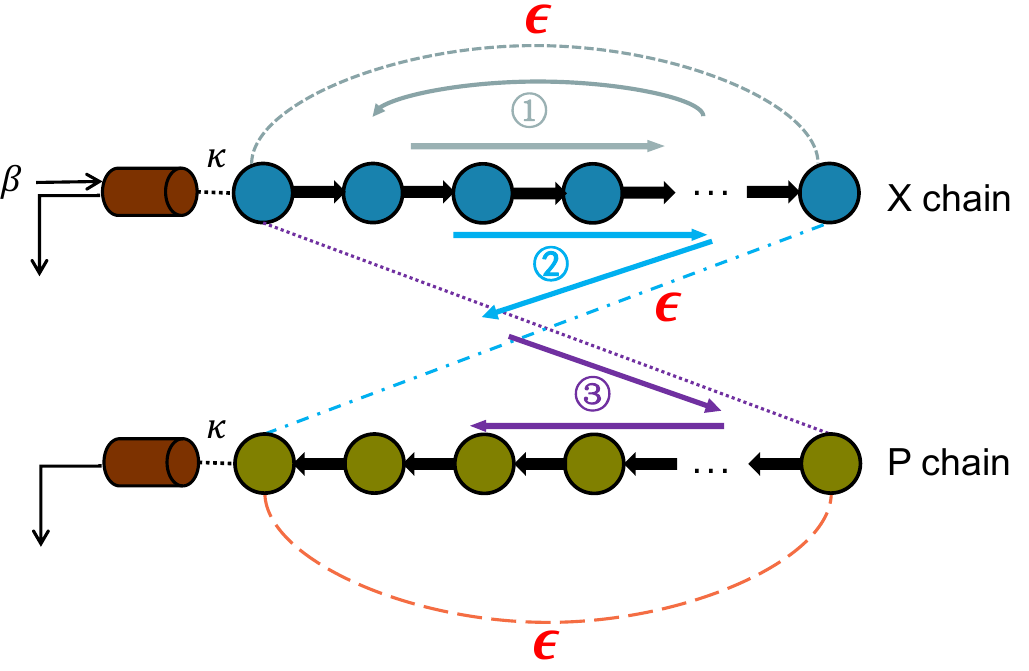} \label{3}
}
\quad
\subfigure[$~\theta=0,~m=N,~\phi \in {[{0,\frac{\pi}{2}})}:~~~~~~~~~~~~~~~$ $\mathcal{S}_\tau(N,\epsilon)\propto e^{2A{(N-1)}},~\bar{n}_{\textsf{tot}}\propto O(1)$, \leftline{$~~~~~~~~~~~~~~\overline{\textrm{SNR}}_\textsf{D}(N)\propto e^{2A(N-1)}$.}]{
\includegraphics[scale=0.48]{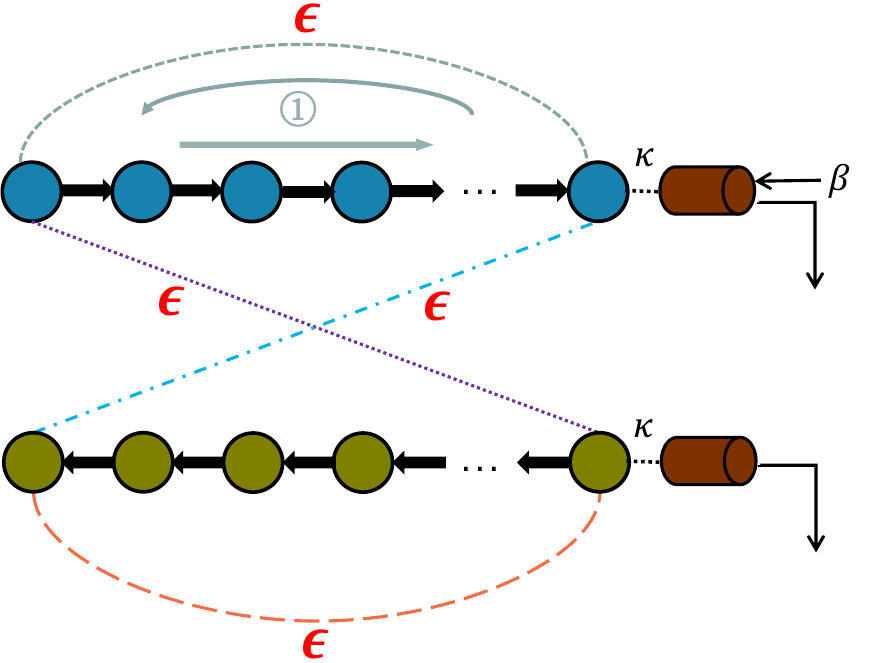} \label{4}
}
\quad
\subfigure[$~\theta=\frac{\pi}{2},~m=1~,~\phi \in {({0,\frac{\pi}{2}}]}:~~~~~~~~~~$
~~~~~~~$~~~\mathcal{S}_\tau(N,\epsilon)\propto e^{2A{(N-1)}},~\bar{n}_{\textsf{tot}}\propto O(1)$, ~~~~~~~~~~~~~~~~~~~\leftline{~~~~~~~~~~~~~~~~~~~~~~$\overline{\textrm{SNR}}_\textsf{D}(N)\propto e^{2A(N-1)}$.}]{
\includegraphics[scale=0.48]{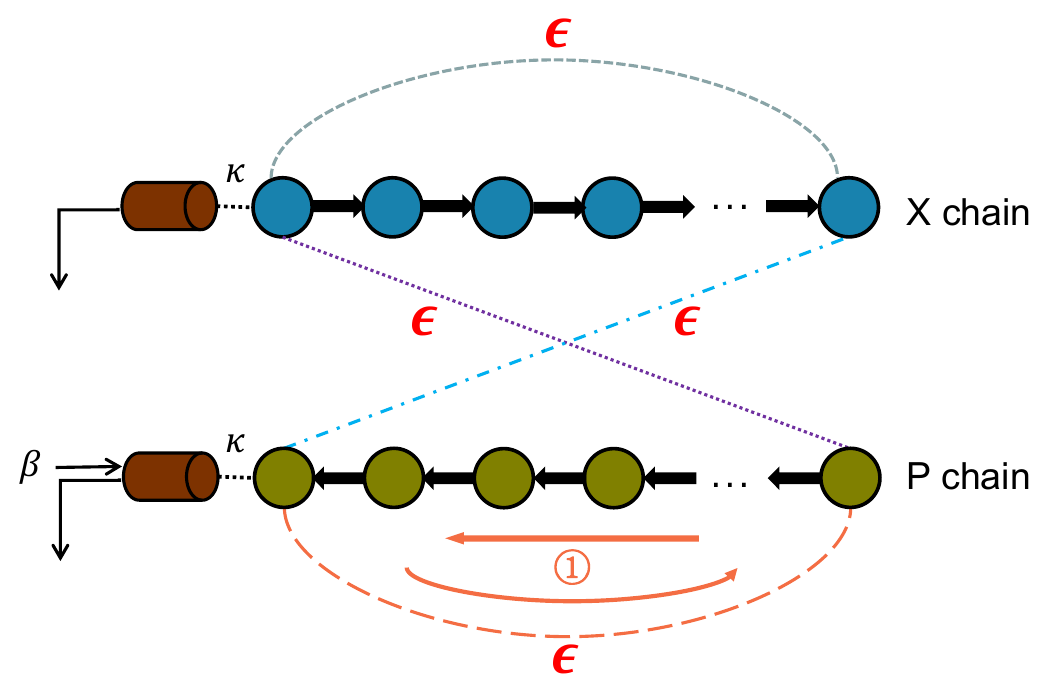} \label{5}
}
\quad
\subfigure[$~\theta=\frac{\pi}{2},~m=N,~\phi \in {[{0,\frac{\pi}{2}}]}:~~~~~~~~~~~~~~~$ ~~~$~~~~~\mathcal{S}_\tau(N,\epsilon)\propto e^{2A{(N-1)}},~\bar{n}_{\textsf{tot}}\propto e^{2A{(N-1)}}$, ~\leftline{$~~~~~~~~~~~~~\overline{\textrm{SNR}}_\textsf{D}(N)\propto O(1).$}]{
\includegraphics[scale=0.48]{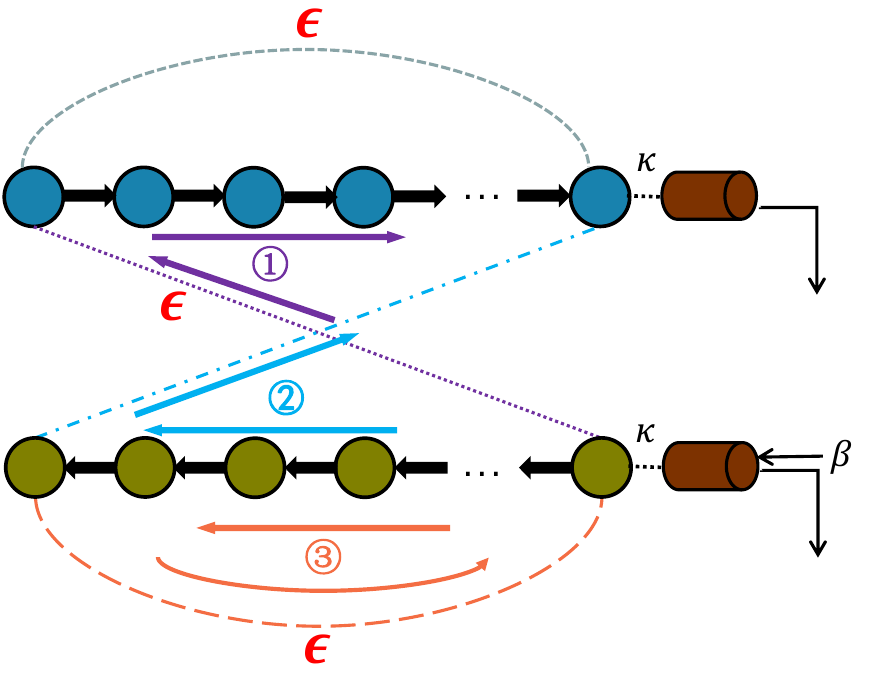} \label{6}
}
\caption{$\overline{\textrm{SNR}}_\textsf{D}$ with different values of $\theta$, $m$ and $\phi$ for $\hat{V}_{\textsf{NHSE}}$. The paths amplifying the signal power are illustrated in each case.}
\end{figure*}

First, we consider the case where a real drive $\beta$, i.e., $\theta=0$, is injected into the chain at some site $m$. From Eq.~(\ref{xpv1}),  if  the perturbation $\epsilon=0$, a real drive only excites the X chain, which is illustrated in Fig.~2(a) and Fig.~2(b).  It can be calculated (see Appendix C) that
\begin{equation}\label{sn1}
\begin{aligned}
\mathcal{S}_\tau(N,\epsilon)=&2\kappa\tau\kappa|\beta|^2\Big{|}\textrm{Re}\big{\{}e^{-i\phi}
\big{[}(h^{-1}_{1,m}\epsilon\textrm{sin}\varphi h^{-1}_{m,N}e^{-A(N-1)}\\
&-h^{-1}_{m,1}\epsilon\textrm{sin}\varphi h^{-1}_{N,m}e^{A(N-1)})+i(h^{-1}_{m,1}\epsilon\textrm{cos}\varphi h^{-1}_{N,m}\\
&+h^{-1}_{1,m}\epsilon\textrm{cos}\varphi h^{-1}_{m,N})e^{-A(2m-N-1)}\big{]}\big{\}}\Big{|}^2,\\
\mathcal{N}_\tau(N,0)=&\frac{1}{2},\\
\bar{n}_{\textsf{tot,D}}(0)=&\kappa|\beta|^2|h^{-1}_{N,m}|^2e^{2A(N-m)},
\end{aligned}
\end{equation}
where $h^{-1}_{m,1}=h^{-1}_{m,N}=h^{-1}_{N,m}=h^{-1}_{1,m}=-\frac{2}{\kappa}$ (see Appendix E).

Suppose that the drive is injected into the chain at site $m=1$, which is the case in \cite{McDonald2020}. From Eq.~(\ref{sn1}) it can be seen that there are three paths amplifying the amplitude of the wavepacket, which have been illustrated in Fig. 2(a), and the  signal power $\mathcal{S}_\tau(N,\epsilon)$
is at most in the order of $e^{2A(N-1)}$. To obtain the SNR per photon, another key factor is the total average number of photons.
Note that for the $\hat{x}$ quadrature excitation at $m=1$, the average number of photons at each site amplifies along the propagation of the wavepacket to the right, and for large $A$  the average photon number at site $N$ is exponentially larger than those on other sites.  In this case  $\bar{n}_{\textsf{tot,D}}(0)$ depicts the average photon number at  site $N$ and satisfies  $\bar{n}_{\textsf{tot}}=\bar{n}_{\textsf{tot,D}}(1+O(e^{-2A}))$. Moreover, from Eq.~(\ref{sn1}),  $\bar{n}_{\textsf{tot,D}}(0)\propto e^{2A(N-1)}$.  This yields the SNR per photon being at most $O(1)$. Thus, as stated in \cite{McDonald2020}, in this case ($\theta=0$ and $m=1$) the NHSE does not provide any true advantage in sensing.

Now we turn to the case where $m=N$.  For the signal power $\mathcal{S}_\tau(N,\epsilon)$, from Eq.~(\ref{sn1}), the signal power $\mathcal{S}_\tau(N,\epsilon)$ is in the order of $e^{2A(N-1)}$ as long as the phase $\varphi$ in $\hat{V}_{\textsf{NHSE}}$ is nonzero and the measurement angle $\phi\neq\frac{\pi}{2}$.  As illustrated in Fig.~2(b), in this situation there is only one path amplifying the amplitude of the wavepacket: for non-zero $\epsilon$, a wavepacket can tunnel from the site $N$  to the site $1$ of the X chain, and then propagates along the X chain back to site $N$ amplifying the signal. Thus to obtain an exponential enhancement of the signal power, at least some portion of the $\hat{x}$  quadrature has to be detected. If the measurement angle $\phi=\frac{\pi}{2}$, then from Eq.~(\ref{M(N)}), only the  $\hat{p}$  quadrature is detected, and the signal power will be exponentially deamplified from Eq.~(\ref{sn1}). For the total average photon number, from Eq.~(\ref{sn1}), $\bar{n}_{\textsf{tot,D}}(0)=\kappa|\beta|^2|h^{-1}_{N,N}|^2=O(1)$. This is because that the $\hat{x}$ quadrature is initially excited at site $N$, and as the wavepacket propagates to the left, the average number of photons at each site is deamplified. Thus, in this case  $\bar{n}_{\textsf{tot,D}}(0)$ depicts the average photon number at site $N$. This leads to
\begin{equation}
\begin{aligned}
\overline{\textrm{SNR}}_\textsf{D}&=16\kappa\tau\Big{|}\textrm{Re} \Big{\{}e^{-i\phi}\big{[}\frac{\epsilon}{\kappa}\textrm{sin}\varphi(e^{-A(N-1)}-e^{A(N-1)})\\
&~~~+i\frac{2\epsilon}{\kappa}\textrm{cos}\varphi e^{-A(N-1)}\big{]}\Big{\}}\Big{|}^2\\
&\propto e^{2A(N-1)},
\end{aligned}
\end{equation}
which means an \textit{exponential enhancement } of the SNR per photon  with the NHSE perturbation Hamiltonian.

Note that in conventional quantum sensing, $N$ sensors interact independently and linearly with the parameter to be measured. The QFI scales at best in the order of $N^2$, which typically requires entanglement of the $N$ sensors. The exponential enhancement of the SNR per photon in our setting is mainly due to two key factors. One is that the Hatano-Nelson chain provides a potentially exponential amplification of the amplitude of the wavepacket with chirality. This unusual non-linear and non-reciprocal amplification makes that the exponential $N$ dependence of the QFI does not violate the standard Heisenberg-limit constrains. The other key factor is to find a proper path along which the signal is amplified exponentially while the total number of photons is not. It is worth noting that the enhancement of the SNR per photon has nothing to do with the exceptional point property.

Now we investigate the case where  the drive is $i|\beta|$, i.e.,  $\theta=\frac{\pi}{2}$.  From Eq. (\ref{xpv1}),  if  the perturbation $\epsilon=0$, the drive only excites the P chain as illustrated in Fig.~2(c) and Fig.~2(d).  In this case, we have (see Appendix C)
\begin{equation}\label{sn2}
\begin{aligned}
\mathcal{S}_\tau(N,\epsilon)=&2\kappa\tau\kappa|\beta|^2\Big{|}
\textrm{Re}\big{\{}e^{-i\phi}\big{[}(h^{-1}_{m,1}\epsilon\textrm{cos}\varphi h ^{-1}_{N,m}\\
&+h^{-1}_{1,m}\epsilon\textrm{cos}\varphi h^{-1}_{m,N})e^{A(2m-N-1)}\\ &+i(h^{-1}_{m,1}\epsilon\textrm{sin}\varphi h^{-1}_{N,m}e^{-A(N-1)}\\
&+h^{-1}_{1,m}\epsilon\textrm{sin}\varphi h^{-1}_{m,N}e^{A(N-1)})\big{]}\big{\}}\Big{|}^2,\\
\mathcal{N}_\tau(N,0)=&\frac{1}{2},\\
\bar{n}_{\textsf{tot,D}}(0)=&\kappa|\beta|^2|h^{-1}_{1,m}|^2e^{2A(m-1)}.
\end{aligned}
\end{equation}

Suppose that the input site is $m=1$. From Eq.~(\ref{sn2}),  the signal power $\mathcal{S}_\tau(N,\epsilon)$ is in the order of $e^{2A(N-1)}$ as long as the phase $\varphi$ in $\hat{V}_{\textsf{NHSE}}$  is nonzero and the measurement angle $\phi\neq0$. As illustrated in
Fig.~2 (c), there is also only one path amplifying the signal power: for non-zero $\epsilon$, a wavepacket can tunnel from the site $1$  to the site $N$ of the P chain, and then propagates leftwards along the P chain back to the site $1$ amplifying the signal. Therefore, to obtain an exponential enhancement of the signal power, at least some portion of the $\hat{p}$  quadrature has to be measured. However, detecting with the measurement angle $\phi=0$ implies that only the $\hat{x}$  quadrature is detected (see Eq.~(\ref{M(N)})). This results in an exponentially reduced signal power from Eq.~(\ref{sn1}). For the total average photon number, from Eq.~(\ref{sn2}), $\bar{n}_{\textsf{tot,D}}(0)=\kappa|\beta|^2|h^{-1}_{N,1}|^2=O(1)$. This is because that the $\hat{p}$ quadrature is  initially excited at the site $1$, and as the wavepacket propagates rightwards, the average number of photons at each site is deamplified. Thus, in this case  $\bar{n}_{\textsf{tot,D}}(0)$ depicts the average photon number at  the site 1. This leads to
\begin{equation}
\begin{aligned}
\overline{\textrm{SNR}}_\textsf{D}&=16\kappa\tau\Big{|}\textrm{Re} \Big{\{}e^{-i\phi}\big{[}\frac{2\epsilon}{\kappa}\textrm{cos}\varphi e^{-A(N-1)}\\
&~~~+i\frac{\epsilon}{\kappa}\textrm{sin}\varphi (e^{-A(N-1)}+e^{A(N-1)})\big{]}\Big{\}}\Big{|}^2\\
&\propto e^{2A(N-1)},
\end{aligned}
\end{equation}
which implies an \textit{exponential enhancement } of the SNR per photon  with the NHSE perturbation Hamiltonian.

For the case where $\theta=\frac{\pi}{2}$ and $m=N$, following a similar analysis it can be seen that  the signal power $\mathcal{S}_\tau(N,\epsilon)$ is in the order of $e^{2A(N-1)}$, while $\bar{n}_{\textsf{tot,D}}(0)\propto e^{2A(N-1)}$.  This yields the SNR per photon being $O(1)$. The corresponding paths amplifying the signal are illustrated in Fig.~2(d).

 \subsection{$\overline{\textrm{SNR}}_{\tau}(N)$ with $\hat{V}_{N}$ }

In this subsection, we consider another perturbation Hamiltonian $\hat{V}_{N}=\hat{a}_N^\dagger\hat{a}_N$. A small $\epsilon$  corresponds to a small detuning of the $N$ site. As illustrated in Fig.~3, for nonzero $\epsilon$, a wavepacket can scatter off the boundary of the X (P) chain and change its quadrature.

\begin{figure*}[htb]
%\centering
\subfigure[$~\theta=0,~m=1,~\phi \in {({0,\frac{\pi}{2}}]}:~~~~~~~~~~~~$ ~~~~~~~~~~~~~~~~~~~~~~$~~~~~~~~~~\mathcal{S}_\tau(N,\epsilon)\propto e^{4A{(N-1)}},~\bar{n}_{\textsf{tot}}\propto e^{2A{(N-1)}}$, \leftline{~~~~~~~~~~~~~~~~~~~~~$\overline{\textrm{SNR}}_\textsf{D}(N)\propto e^{2A{(N-1)}}$.} ]{
\includegraphics[scale=0.45]{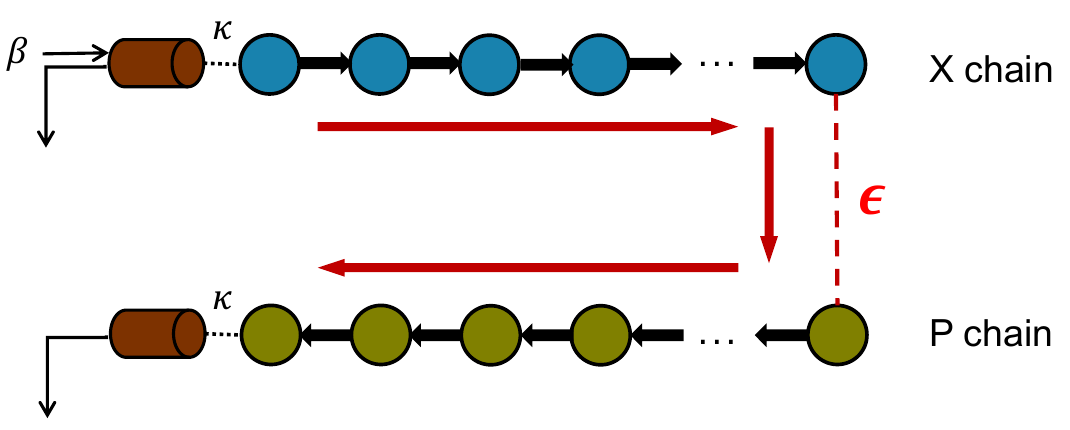} \label{3}
}
\quad
\subfigure[$~\theta=0,~m=N,~\phi \in {({0,\frac{\pi}{2}}]}:$ ~~~~~~~~~~~~~~~~~~~~~~~~~~~ $~~~~~~~~~~\mathcal{S}_\tau(N,\epsilon)\propto O(1),~\bar{n}_{\textsf{tot}}\propto O(1)$, \leftline{~~~~~~~~~~~~~~~~~~~~$\overline{\textrm{SNR}}_\textsf{D}(N)\propto O(1)$.}]{
\includegraphics[scale=0.45]{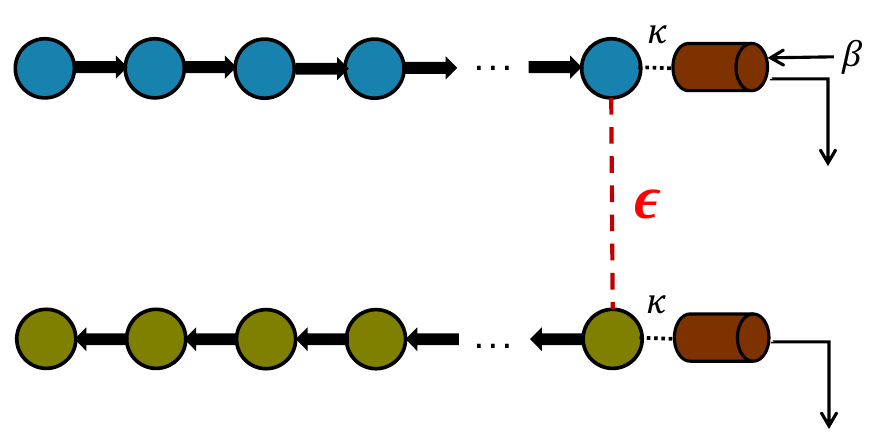} \label{4}
}
\quad
\subfigure[$~\theta=\frac{\pi}{2},~m=1,~\phi \in {[{0,\frac{\pi}{2}})}:~~~~~~~~~~~~$ $~~~~\mathcal{S}_\tau(N,\epsilon)\propto e^{-4A{(N-1)}},~\bar{n}_{\textsf{tot}}\propto O(1)$, \leftline{~~~~~~~~~~~~~~~~~~~~$\overline{\textrm{SNR}}_\textsf{D}(N)\propto e^{-4A(N-1)}$.}]{
\includegraphics[scale=0.45]{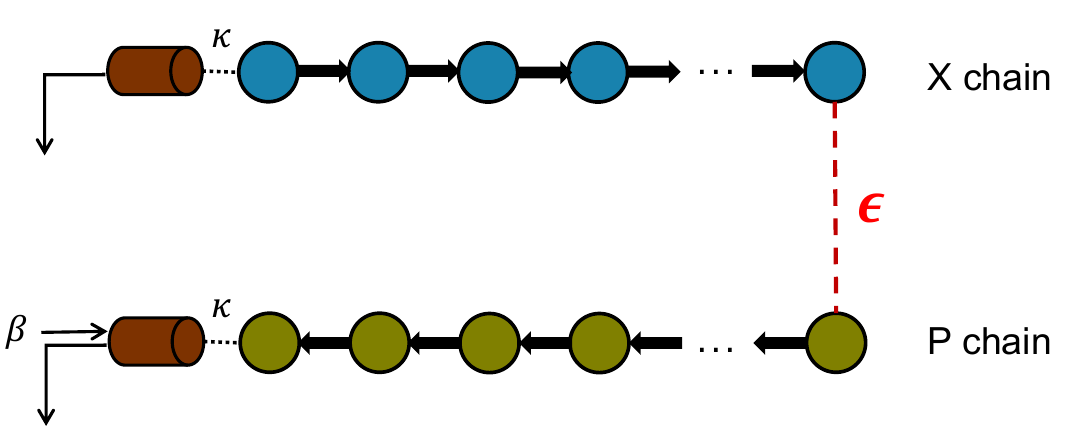} \label{5}
}
\quad
\subfigure[$~\theta=\frac{\pi}{2},~m=N,~\phi \in {[{0,\frac{\pi}{2}})}:$ \leftline{~~~~~~~~~~~~~~~~~~~$\mathcal{S}_\tau(N,\epsilon)\propto O(1),~\bar{n}_{\textsf{tot}}\propto e^{2A{(N-1)}}$,} \leftline{~~~~~~~~~~~~~~~~~~~$\overline{\textrm{SNR}}_\textsf{D}(N)\propto e^{-2A{(N-1)}}$.}]{
\includegraphics[scale=0.45]{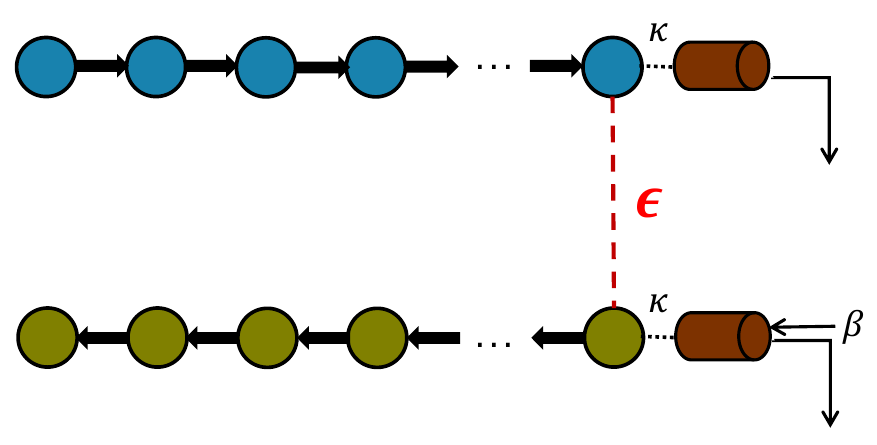} \label{6}
}
\caption{$\overline{\textrm{SNR}}_D$  with different values of $\theta$, $m$ and $\phi$ for $\hat{V}_{N}$. The path amplifying the signal power is illustrated in each case.}
\end{figure*}

In \cite{McDonald2020}, it was demonstrated that $\hat{V}_{N}=\hat{a}^\dagger_{N}\hat{a}_{N}$ can provide an exponential enhancement of the SNR per photon in the case of $\theta=0$ and $m=1$. Now we compute the SNR per photon with different parameters of the coherent drive.
Similar to the case in Sec. III A, we focus on the dominate term $\overline{\textrm{SNR}}_\textsf{D}$ of $\overline{\textrm{SNR}}_{\tau}(N)$, since the following relationship holds
\begin{equation}
\overline{\textrm{SNR}}_{\tau}(N)=\overline{\textrm{SNR}}_\textsf{D}(N)\cdot\Big{(}1+O(e^{-2A})\Big{)}.
\end{equation}
As shown in Appendix D, the $\overline{\textrm{SNR}}_\textsf{D}$  is
\begin{equation}
\begin{aligned}
&\overline{\textrm{SNR}}_\textsf{D}\\
=&4\kappa\tau\big{|}h^{-1}_{m,N}h^{-1}_{N,m}\big{|}^2\\
&\cdot\frac{\big{|}-\epsilon\textrm{sin}\theta\textrm{cos}\phi e^{-2A(N-m)}+\epsilon\textrm{cos}\theta\textrm{sin}\phi e^{2A(N-m)}\big{|}^2}{|h^{-1}_{N,m}|^2\textrm{cos}^2\theta e^{2A(N-m)}+|h^{-1}_{1,m}|^2\textrm{sin}^2\theta e^{2A(m-1)}},
\end{aligned}
\end{equation}
which is continuously differentiable with respect to parameters ($\theta$,~$\phi$). It is straightforward to obtain the extrema at $(\theta,~\phi)=$ $(0,~\frac{\pi}{2})$ or $(\frac{\pi}{2},~0)$. The corresponding extreme value is
\begin{equation}
\begin{aligned}
\overline{\textrm{SNR}}_\textsf{D}|_{\theta=0,~\phi=\frac{\pi}{2}}&=4\kappa\tau\epsilon^2|h^{-1}_{m,N}|^2e^{2A(N-m)},\\
\overline{\textrm{SNR}}_\textsf{D}|_{\theta=\frac{\pi}{2},~\phi=0}&=4\kappa\tau\epsilon^2|h^{-1}_{m,N}|^2e^{2A(-2N+m+1)}.\\
\end{aligned}
\end{equation}

When $\theta=0$, $m=1$ and $\phi=\frac{\pi}{2}$, $\overline{\textrm{SNR}}_\textsf{D} \propto e^{2A(N-1)}$ takes its  maximum. This corresponds to the case in \cite{McDonald2020}, where a real drive $\beta$ is injected into the chain at site 1 to excite the X chain and the $\hat{p}$ quadrature at site 1 is detected. The corresponding path amplifying the signal power is illustrated in Fig.~3(a).

If $\theta=0$ and $m=N$, then $\overline{\textrm{SNR}}_\textsf{D} =O(1)$. This is because as illustrated in Fig.~3(b), if a real drive is injected into the chain at site $N$, after two traversals (leftwards and rightwards) of the X (P) chain, the net amplification factor of the signal power is $O(1)$, while the average photon number is dominated by the site $N$ with $O(1)$.

Let us turn to the case where the  drive is $i|\beta|$, i.e., $\theta=\frac{\pi}{2}$, corresponding to the $\hat{p}$ quadrature excitation. If $m=1$ (Fig.~3(c)), then the signal power is deamplified as  the wavepacket propagates rightwards along the P chain. After it scatters off to the X chain, it is still deamplified as it goes back to the site 1 along the X chain. This yields the signal power in the order of  $e^{-4A(N-1)}$. The total average photon number is dominated by that at the site $1$, which is $O(1)$. Thus, in this situation, the SNR per photon is   \textit{exponentially deamplified} as $O(e^{-4A(N-1)})$. If $m=N$ (Fig.~3(d)),  the net amplification factor of the signal power is $O(1)$ after two traversals (leftwards and rightwards) of the X (P) chain, while the average photon number is dominated by that at  site $1$, which is amplified by a factor of $O(e^{2A(N-1)})$. This yields the SNR per photon with the order of $O(e^{-2A(N-1)})$, which also implies an\textit{ exponentially reduction} of sensing.

Now it is clear that  when utilizing a non-Hermitian sensor, it is important to  optimize its performance  by choosing appropriate parameters. Otherwise,  a sensor, which has the potential to provide an exponential enhancement in sensing, could only extract exponentially deamplified signal.

\section{Beyond linear response}

Up to now, we have assumed that the parameter $\epsilon$ to be sensed is infinitesimal, and thus we only need to compute  the first order of the signal in $\epsilon$, i.e., the linear response  of the output field. In this section, we consider the case where the parameter to be measured $\epsilon_0$ may not be very small such that all orders in $\epsilon_0$   of the output fields have to be calculated. We focus on the most interesting case where $\epsilon_0/\kappa\ll 1$, while $e^{A(N-1)}\epsilon_0/\kappa$ is not necessarily small for large $A$, as in \cite{McDonald2020}.

Since the system is still  Gaussian  for any $\epsilon_0$, we can still use the SNR per photon as a figure of merit. However, in contrast to the case where $\epsilon$  is infinitesimal, we have to take account of the effect of $\epsilon_0$ in the noise power and the total average photon number. The definition of the signal-to-noise ratio becomes:
\begin{equation}\label{SNRBeyond}
\begin{aligned}
\textrm{SNR}_\tau(N,\epsilon_0)\equiv\frac{|\langle\hat{\mathcal{M}}_\tau(N)\rangle_{\epsilon_0}
-\langle\hat{\mathcal{M}}_\tau(N)\rangle_0|^2}{\frac{\mathcal{N}_\tau(N,0)+\mathcal{N}_\tau(N,\epsilon_0)}{2}}.
\end{aligned}
\end{equation}
As stated in \cite{McDonald2020}, $\textrm{SNR}_\tau(N,\epsilon_0)$ quantifies the distinguishability between the Gaussian Homodyne current distributions under parameter $\epsilon=0$ and $\epsilon=\epsilon_0$.
The corresponding total average number of photons is
\begin{equation}\label{photonbeyond}
\begin{aligned}
\bar{n}_{\textsf{tot}}\equiv\frac{\bar{n}_{\textsf{tot}}(0)+\bar{n}_{\textsf{tot}}(\epsilon_0)}{2},
\end{aligned}
\end{equation}
and $\overline{\textrm{SNR}}_{\tau}(N,\epsilon_0)=\textrm{SNR}_\tau(N,\epsilon_0)/\bar{n}_{\textsf{tot}}$.

For the local perturbation Hamiltonian $\hat{V}_N=\hat{a}_N^\dagger\hat{a}_N$, it was demonstrated in \cite{McDonald2020} that increasing the amplification factor $A$ and/or the size $N$ indefinitely is no longer optimal. Actually, in the $\epsilon_0\ll \kappa$ regime, to obtain an optimized amplification of the SNR per photon, the optimal amplification factor $A^*$ and the size $N^*$ should satisfy \cite{McDonald2020}
\begin{equation}
e^{A^*(N^*-1)}\approx\sqrt[4]{\frac{\kappa^2}{8\epsilon^2_0}},
\end{equation}
and the optimized SNR per photon for large $A^*$ is
\begin{equation}\label{vn}
\overline{\textrm{SNR}}_{\tau}(N^*,\epsilon_0)=2\sqrt{2}\kappa \tau \frac{\epsilon_0}{\kappa}(1-O(e^{-4A^*})).
\end{equation}

We now focus on the NHSE perturbation Hamiltonian $\hat{V}_{\textsf{NHSE}}=e^{i\varphi}\hat{a}^\dagger _1\hat{a}_N+e^{-i\varphi}\hat{a}_1\hat{a}^\dagger_N$, and demonstrate that in this situation the SNR per photon can be enhanced at least by a large factor ${\kappa}/{\epsilon_0} $ over that obtained by using $\hat{V}_N=\hat{a}_N^\dagger\hat{a}_N$ as  Eq. (\ref{vn}).

For simplicity, assume that  $\theta=0$, $m=N$ and the measured phase $\phi=0$.  This corresponds to $\hat{x}$ quadrature excitation by a real drive at site $N$. It can be computed that the signal power
\begin{equation}\label{signalbeyond}
\begin{aligned}
\mathcal{S}_\tau(N,\epsilon_0)=2\kappa\tau \kappa|\beta|^2\Big{ |}\tilde{\mathbb{H}}[\epsilon_0]^{-1}_{N,N}-\tilde{\mathbb{H}}[0]^{-1}_{N,N}\Big{|}^2,
\end{aligned}
\end{equation}
where $\tilde{\mathbb{H}}[\epsilon_0]^{-1}$ is dependent on $\epsilon_0$ and $e^{A(N-1)}$  (see Appendix E and F).

Taking the noise of the input field to be vacuum ($\bar{n}_{\textsf{th}}=0$), we have
\begin{equation}\label{noisebeyond}
\begin{aligned}
\mathcal{N}_\tau&=\frac{\mathcal{N}_\tau(N,0)+\mathcal{N}_\tau(N,\epsilon_0)}{2}\\
&=\frac{1}{4}\Big(1+(1+\kappa\mathbb{\tilde{H}}[\epsilon_0]^{-1}_{N,N})^2\\
&~~~+\kappa^2
[\mathbb{\tilde{H}}[\epsilon_0]^{-1}_{N,2N}]^2e^{4A(N-1)}\Big).
\end{aligned}
\end{equation}

The total number of photons is
\begin{equation}\label{numberbeyond}
\begin{aligned}
\bar{n}_{\textsf{tot}}=&\frac{1}{2}\kappa |\beta|^2 \bigg\{\sum^N_{n=1}[\mathbb{\tilde{H}}[0]^{-1}_{n,N}]^2e^{-2A(N-n)}\\
&+\sum^N_{n=1}\Big([\mathbb{\tilde{H}}[\epsilon_0]^{-1}_{n,N}]^2e^{-2A(N-n)}\\
&+[\mathbb{\tilde{H}}[\epsilon_0]^{-1}_{N+n,N}]^2
e^{-2A(N+n-2)}\Big)\bigg\}.
\end{aligned}
\end{equation}
Let
\begin{equation}\label{numb1}
\begin{aligned}
\bar{n}_{\textsf{tot,D}}=&\frac{1}{2}\kappa |\beta|^2\Big\{[\mathbb{\tilde{H}}[0]^{-1}_{N,N}]^2+[\mathbb{\tilde{H}}[\epsilon_0]^{-1}_{N,N}]^2\\
&+[\mathbb{\tilde{H}}[\epsilon_0]^{-1}_{N+1,N}]^2e^{-2A(N-1)}   \Big\}.
\end{aligned}
\end{equation}
It is clear that $\bar{n}_{\textsf{tot}}=\bar{n}_{\textsf{tot,D}}\cdot Q(A)$, where $Q(A)=1+O(e^{-2A})$. Thus, $\bar{n}_{\textsf{tot,D}}$ describes the dominate term of $\bar{n}_{\textsf{tot}}$ for large $A$.

For further analysis, we take $\varphi=\frac{\pi}{2}$.  According to Eqs.~\eqref{SNRBeyond}, \eqref{signalbeyond}, \eqref{noisebeyond} and \eqref{numb1}, we have
\begin{equation}
\begin{aligned}
 \overline{\textrm{SNR}}_{\textsf{D}}(N,\epsilon_0)\equiv\frac{\mathcal{S}_\tau(N,\epsilon_0)}
 {\mathcal{N}_\tau(N,\epsilon_0)\cdot\bar{n}_{\textsf{tot,D}}}= \kappa \tau \frac{G_1(\epsilon_0, \eta)}{G_2(\epsilon_0, \eta)},
\end{aligned}
\end{equation}
where $\eta\equiv e^{A(N-1)}$ and
\begin{equation}
\begin{aligned}
G_1(\epsilon_0, \eta)
=&16 \epsilon^2_0(1-\eta^2)^2(\eta\kappa+2\epsilon_0-2\eta^2\epsilon_0)^2,\\
G_2(\epsilon_0, \eta)=&\big(4\epsilon^2_0+4\eta^4\epsilon^2_0+\eta^2(\kappa^2-8\epsilon^2_0)\big)\\
\cdot&\big(2\eta\kappa\epsilon_0-2\eta^3\kappa\epsilon_0+2\epsilon^2_0+2\eta^4\epsilon^2_0\\
&+\eta^2(\kappa^2-4\epsilon^2)\big).
\end{aligned}
\end{equation}
It is clear that $\frac{G_1(\epsilon_0, \eta)}{G_2(\epsilon_0, \eta)}$ does not necessarily increase as $A$ and/or $N$ increase as in the linear response case. Thus we need to optimize $\eta$.

If $e^{A(N-1)}=\sqrt[4]{\frac{\kappa^2}{8\epsilon^2_0}}$, the SNR per photon with perturbation $\hat{V}_{\textsf{NHSE}}=e^{i\varphi}\hat{a}^\dagger _1\hat{a}_N+e^{-i\varphi}\hat{a}_1\hat{a}^\dagger_N$ is
\begin{equation}
\begin{aligned}
 \overline{\textrm{SNR}}_{\textsf{D}}(N,\epsilon_0)\Bigg|_{e^{A(N-1)}=\sqrt[4]{\frac{\kappa^2}{8\epsilon^2_0}}}= 4\sqrt{2}\kappa\tau~\frac{\epsilon_0}{\kappa}+O(\frac{\epsilon_0}{\kappa})^{\frac{3}{2}},
\end{aligned}
\end{equation}
which is in the same order as the optimized amplification utilizing $\hat{V}_N=\hat{a}_N^\dagger\hat{a}_N$  in \cite{McDonald2020} (see  Eq. (\ref{vn})).

If $e^{A(N-1)}=\frac{\kappa}{\epsilon_0}$, we have
\begin{equation}
\begin{aligned}
 \overline{\textrm{SNR}}_{\textsf{D}}(N,\epsilon_0)\Bigg|_{e^{A(N-1)}=\frac{\kappa}{\epsilon_0}}= \frac{16}{5}\kappa\tau+O(\frac{\epsilon_0}{\kappa})^2,
\end{aligned}
\end{equation}
which  increases by a large factor  $O(\kappa/\epsilon_0)$ over that in  Eq.~(\ref{vn}). This implies that in the regime where the linear response is invalid, the non-Hermitian skin effect can provide a much higher SNR per photon than that obtained under the perturbation Hamiltonian $\hat{V}_N=\hat{a}_N^\dagger\hat{a}_N$.

%\begin{figure}[htbp]
%\centering
%\addtocounter{figure}{-1}
%\subfigure{
%\begin{minipage}[t]{1\linewidth}
%\includegraphics[scale=0.6]{beyond_1}
%\caption{$~\overline{\textrm{SNR}}_{\textsf{D}}/{\kappa\tau}$ versus $e^{A(N-1)}$ with the local perturbation Hamiltonian $\hat{V}_N$.~~~~~~~~~}
%\end{minipage}
%}
%\quad
%\subfigure{
%\begin{minipage}[t]{1\linewidth}
%\includegraphics[scale=0.6]{beyond_2}
%\caption{$~\overline{\textrm{SNR}}_{\textsf{D}}/{\kappa\tau}$~versus~ $e^{A(N-1)}$~with~the~perturbation
%~~~~Hamiltonian~$\hat{V}_{\textsf{NHSE}}$.}
%\end{minipage}
%}
%\caption{The dominate term of the SNR per photon versus $e^{A(N-1)}$ with different perturbation Hamiltonians.}
%\end{figure}

\begin{figure}[htbp]
\centering
\subfigure[$~\overline{\textrm{SNR}}_{\textsf{D}}/{\kappa\tau}$ versus $e^{A(N-1)}$ with the local perturbation \leftline{~~~~~~~~~~~~~Hamiltonian $\hat{V}_N$.}]{
\includegraphics[scale=0.55]{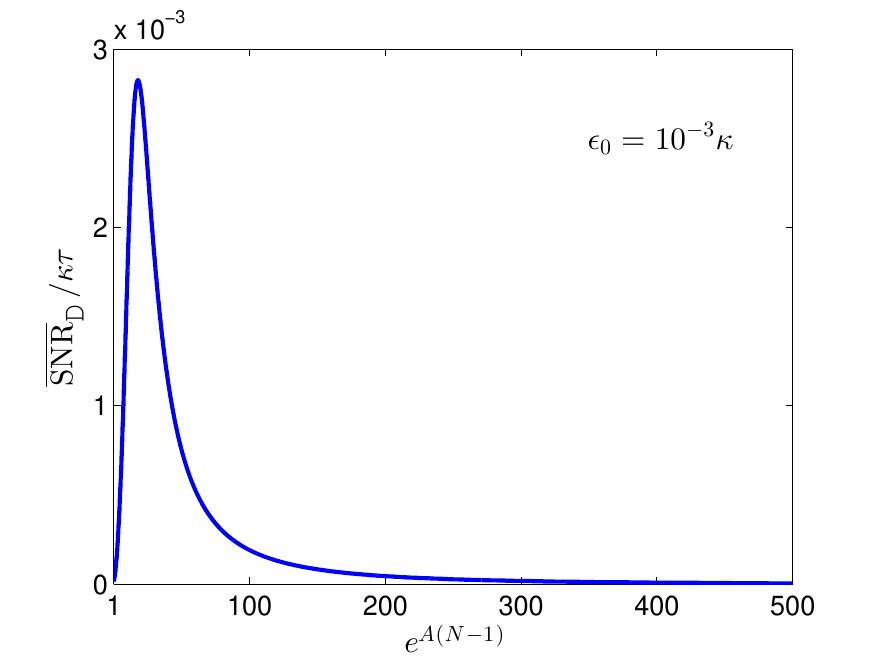}
}
\quad
\subfigure[$~\overline{\textrm{SNR}}_{\textsf{D}}/{\kappa\tau}$~versus~ $e^{A(N-1)}$~with~the~perturbation
\leftline{~~~~~~~~~~~~~~~~Hamiltonian~$\hat{V}_{\textsf{NHSE}}$.}]{
\includegraphics[scale=0.55]{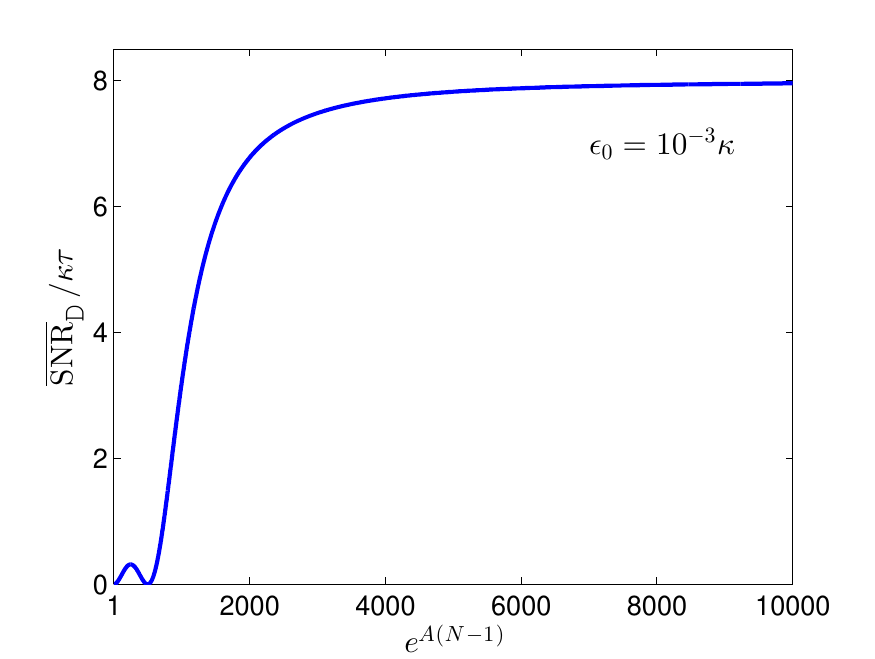}
}
\caption{The dominate term of the SNR per photon versus $e^{A(N-1)}$ with different perturbation Hamiltonians.}
\end{figure}

Note that in obtaining the dominate term of the SNR per photon, a large amplification factor $A$ has been assumed both in the cases of local perturbation and NHSE perturbation.  Moreover, recall that to obtain the optimal SNR per photon under $\hat{V}_N=\hat{a}_N^\dagger\hat{a}_N$,  the optimal amplification factor $A^*$ and the size $N^*$ should satisfy
$e^{A^*(N^*-1)}\approx\sqrt[4]{\frac{\kappa^2}{8\epsilon^2_0}}.$  We illustrate the dominate term of the SNR per photon using $\hat{V}_N$ as a function of $e^{A(N-1)}$  in Fig.~4(a). It can be seen that  the performance of $\overline{\textrm{SNR}}_{\textsf{D}}$ decays quickly as $e^{A(N-1)}$ deviates from the optimal value. This will result in some inconvenience in practice. The first one is that to obtain the best SNR per photon, we have to estimate the optimal amplification factor accurately, which in turn depends on the parameter to be measured. Thus, many rounds of adaptive measurement should be performed. In addition, in practice due to the limitation of $\kappa$, the amplification factor $A$ may not be sufficiently large. This will  lead to  Eq.~(\ref{vn}) being invalid.

As for the NHSE perturbation, we illustrate the dominate term  $\overline{\textrm{SNR}}_{\textsf{D}}(N,\epsilon_0)$  of the SNR per photon versus $e^{A(N-1)}$  in Fig.~4(b). It can be seen that as $e^{A(N-1)}$ increases, $\overline{\textrm{SNR}}_{\textsf{D}}$ approaches to a constant value, which is much larger than the maximum $\overline{\textrm{SNR}}_{\textsf{D}}$ in Fig.~4 (a). In contrast to the case using $\hat{V}_N$,  to obtain a high SNR per photon utilizing $\hat{V}_{\textsf{NHSE}}$, we  just need to choose a  sufficiently large  amplification factor $A$. This  not only ensures the validity of the analysis, but also makes that the number of cavities can be as few as three. In this sense we remark that using $\hat{V}_{\textsf{NHSE}}$ for sensing is more robust.

\section{Conclusion and Discussion}
In this paper, we have demonstrated the importance of optimizing the parameters of the coherent drive in non-Hermitian sensing by showing that  the SNR per photon can be exponentially-enhanced or exponentially-reduced depending on parameters of the drive and detection. If the parameter to be sensed is infinitesimal, we have demonstrated that  for large amplification by choosing appropriate parameters of the drive, the NHSE can provide an exponential enhancement of sensing. Moreover, the advantage of using NHSE persists in the regime beyond linear response. Utilizing NHSE, the SNR per photon can be enhanced at least by a large factor of $\kappa/\epsilon_0$ over that obtained by using the local perturbation $\hat{V}_N$, and the protocol is more robust in tuning the amplification factor. Our setup could be realized in a variety of quantum optical platforms and superconducting circuits. An interesting question worth  further exploring is whether the exponential enhancement can still be maintained in the presence of gain and loss quantum noises.

\section*{Acknowledgments}
We thank A. McDonald for helpful discussions. B. Q. acknowledged the support of National Natural Science Foundation of China (Nos.  11688101, 61773370, 61833010 and 61621003), and D. D. acknowledged the support of the Australian Research Council’s Discovery Projects funding scheme under Project DP190101566.

%and D. D. acknowledged the partial support by the Australian Research Councils Discovery Projects funding scheme under Project DP190101566, the Alexander von Humboldt Foundation of Germany and the U.S. Office of Naval Research Global under Grant N62909-19-1-2129. F.N. is supported in part by: NTT Research,
%Japan Science and Technology Agency (JST)
%(via the Q-LEAP program, Moonshot RD Grant No. JPMJMS2061, and the CREST Grant No. JPMJCR1676),
%Japan Society for the Promotion of Science (JSPS) (via the KAKENHI Grant No. JP20H00134 and the JSPS-RFBR Grant No. JPJSBP120194828),
%Army Research Office (ARO) (Grant No. W911NF-18-1-0358),
%Asian Office of Aerospace Research and Development (AOARD) (via Grant No. FA2386-20-1-4069),
%and the Foundational Questions Institute Fund (FQXi) via Grant No. FQXi-IAF19-06.

\clearpage

\begin{widetext}

\section*{Appendices}
To make the paper self-contained, in the Appendices we first introduce the total Hamiltonian of the non-Hermitian sensor in Appendix A. In Appendix B, we present details of the QFI in the Gaussian state and its relation to the SNR. Then  we give the detailed derivations of calculating the SNR per photon. We calculate the SNR per photon with  the NHSE perturbation Hamiltonian and the local perturbation Hamiltonian in Appendix C and Appendix D, respectively. Appendix E presents the derivation of the inverse of $\mathbb{\tilde{H}}[\epsilon]$, and Appendix F calculates all the orders of $\mathbb{\tilde{H}}[\epsilon_0]^{-1}$ with respect to $\epsilon_0$.

\appendix

\section{Effective non-Hermitian system Hamiltonian}

The total Hamiltonian of the sensor is described by
\begin{equation}
\begin{aligned}
\hat{H}_{\text{tot}}=\hat{H}_S+\hat{H}_{\textrm{wave}}+\hat{H}_\textrm{int}+\hat{H}_{\epsilon}+\hat{H}_{\textrm{input}},
\end{aligned}
\end{equation}
where the Hamiltonian of the waveguide is
\begin{equation}
\begin{aligned}
\hat{H}_{\textrm{wave}}=\int d k (k \hat{b}_k^\dagger \hat{b}_k),
\end{aligned}
\end{equation}
the interaction Hamiltonian between the chain and the waveguide is
\begin{equation}
\begin{aligned}
\hat{H}_\textrm{int}=\int dk \frac{1}{\sqrt{\pi}}\sqrt{\frac{\kappa}{2}}(\hat{a}_m \hat{b}_k^\dagger +\hat{a}_m^\dagger \hat{b}_k),
\end{aligned}
\end{equation}
the perturbation Hamiltonian associated with the disturbance $\epsilon$ is
\begin{equation}
\begin{aligned}
\hat{H}_{\epsilon}=\epsilon\hat{V},
\end{aligned}
\end{equation}
and the input Hamiltonian is
\begin{equation}
\begin{aligned}
\hat{H}_{\textrm{input}}=-i\sqrt{\kappa}(\hat{a}_m^\dagger \beta-\hat{a}_m \beta^\dagger).
\end{aligned}
\end{equation}
Here, $\hat{a}_i$ denotes the mode annihilation operator on site $i$ and $\hat{b}_k$ is the annihilation operator of the mode with wave number $k$ in the waveguide which satisfies $\big[\hat{b}_k,~\hat{b}_{k'}^\dagger\big]=\delta(k-k')$.

The Heisenberg equations of motion for the cavity mode and the waveguide are
\begin{equation}\label{abmotion}
\begin{aligned}
\dot{\hat{a}}_n=&w \hat{a}_{n-1}+\Delta\hat{a}^\dagger_{n+1}+\Delta\hat{a}^\dagger_{n-1}-w\hat{a}_{n+1}-i\epsilon[\hat{a}_n,\hat{V}]-\sqrt{\kappa}\beta\delta_{n,m}\\
&-i\delta_{n,m}\int dk (\frac{1}{\sqrt{\pi}}\sqrt{\frac{\kappa}{2}} \hat{b}_k),\\
\dot{\hat{b}}_k=&-ik\hat{b}_k-i\frac{1}{\sqrt{\pi}}\sqrt{\frac{\kappa}{2}} \hat{a}_m,
\end{aligned}
\end{equation}
respectively. The solution of the second equation in Eq. \eqref{abmotion} is \begin{equation}\label{bmotion}
\begin{aligned}
\hat{b}_k=e^{-ik(t-t_0)}\hat{b}_k(t_0)-i\frac{1}{\sqrt{\pi}}\sqrt{\frac{\kappa}{2}}\int_{t_0}^t dt' e^{-i k (t-t')}\hat{a}_m(t').
\end{aligned}
\end{equation}
Substituting Eq. \eqref{bmotion} into the last term of the first equation in  Eq. \eqref{abmotion} yields
\begin{equation}
\begin{aligned}
\dot{\hat{a}}_n=&w \hat{a}_{n-1}+\Delta\hat{a}^\dagger_{n+1}+\Delta\hat{a}^\dagger_{n-1}-w\hat{a}_{n+1}-i\epsilon[\hat{a}_n,\hat{V}]-\sqrt{\kappa}\beta\delta_{n,m}\\
&-i\sqrt{\frac{\kappa}{2\pi}}\delta_{n,m}\int dk e^{-ik(t-t_0)}\hat{b}_k(t_0)-\delta_{n,m}\frac{\kappa}{2\pi}\int dk \int_{t_0}^t dt' e^{-i k (t-t')}\hat{a}_m(t')\\
=&w \hat{a}_{n-1}+\Delta\hat{a}^\dagger_{n+1}+\Delta\hat{a}^\dagger_{n-1}-w\hat{a}_{n+1}-i\epsilon[\hat{a}_n,\hat{V}]
-\frac{\kappa}{2}\hat{a}_m(t)\delta_{n,m}-\sqrt{\kappa}(\hat{B}^{\textsf{in}}+\beta)\delta_{n,m},
\end{aligned}
\end{equation}
where we have defined $\hat{B}^{\textsf{in}}=i\sqrt{\frac{1}{2\pi}}\int dk e^{-ik(t-t_0)}\hat{b}_k(t_0)$, and used the equations $\int dk e^{-ik(t-t')}=2\pi \delta(t-t')$ and $\int_{t_0}^t dt' \delta(t-t')\hat{a}_m(t')=\frac{1}{2}\hat{a}_m(t)$. Therefore, the effective Hamiltonian of the full system can be expressed as Eq. \eqref{fullHamiltonian}, $$\hat{H}[\epsilon]=\hat{H}_S+\epsilon\hat{V}+\hat{H}_\kappa-i\sqrt{\kappa}(\hat{a}_m^\dagger \beta-\hat{a}_{m}\beta^\dagger).$$  More detailed derivations can be found in \cite{Gardiner2000} and \cite{Clerk2010}.

\section{QFI for Gaussian systems and its relation with SNR}

As mentioned in Eq. (\ref{Bout(N)}), the temporal mode of the output field in the limit of long integration time $\tau$ can be described as
\begin{equation}
\begin{aligned}
\hat{\mathcal{B}}_\tau(N)=\frac{1}{\sqrt{\tau}}\int^\tau_0dt \hat{B}^{\textsf{out}}(t),
\end{aligned}
\end{equation}
which is affected by the change in parameter $\epsilon$. We use observation $\hat{\mathcal{M}}_\tau(N)$ as Eq.~\eqref{M(N)} to extract information about the parameter $\epsilon$ in the output field $\hat{\mathcal{B}}_\tau(N)$. The possible measurements $y$ can be described by a probability distribution $P_\epsilon(y)$, which depends on parameter $\epsilon$. For infinitesimal $\epsilon$, the statistical distance between $P_\epsilon(y)$ and $P_0(y)$ can be depicted by $\epsilon^2 \mathcal{F}$, where $\mathcal{F}$ is the Fisher information, and the QFI can be obtained by optimizing all possible measurement angles of $\hat{\mathcal{M}}_\tau(N)$.

Because of the Gaussian nature of our system, QFI can be computed as \cite{Banchi2015,OH2019}
\begin{equation}\label{QFIQFI}
\begin{aligned}
\text{QFI}=\frac{d\mu_\epsilon}{d\epsilon}V^{-1}_\epsilon\frac{d\mu_\epsilon^\top}{d\epsilon}+\frac{1}{2}\textrm{Tr}[V^{-1}_\epsilon\frac{dV_\epsilon}{d\epsilon}V^{-1}_\epsilon\frac{dV_\epsilon}{d\epsilon}],
\end{aligned}
\end{equation}
where $\mu_\epsilon\triangleq(\langle\hat{\mathcal{X}}\rangle_\epsilon,\langle\hat{\mathcal{P}}\rangle_\epsilon)$ and $V_\epsilon\triangleq\frac{1}{2}
\begin{pmatrix}
     \langle\{\delta\hat{\mathcal{X}}, \delta\hat{\mathcal{X}}\}\rangle_\epsilon & \langle\{\delta\hat{\mathcal{X}}, \delta\hat{\mathcal{P}}\}\rangle_\epsilon \\
     \langle\{\delta\hat{\mathcal{P}}, \delta\hat{\mathcal{X}}\}\rangle_\epsilon & \langle\{\delta\hat{\mathcal{P}}, \delta\hat{\mathcal{P}}\}\rangle_\epsilon
\end{pmatrix}$ are the first and second moments of the Gaussian state with $\hat{\mathcal{X}}=\frac{\hat{\mathcal{B}}_\tau+\hat{\mathcal{B}}_\tau^\dagger}{\sqrt{2}}$ and $\hat{\mathcal{P}}=i\frac{-\hat{\mathcal{B}}_\tau+\hat{\mathcal{B}}_\tau^\dagger}{\sqrt{2}}$ being the quadratures of the output field $\hat{\mathcal{B}}_\tau$, and $\delta\hat{\mathcal{X}}\triangleq\hat{\mathcal{X}}-\langle\hat{\mathcal{X}}\rangle_\epsilon$, $\delta\hat{\mathcal{P}}\triangleq\hat{\mathcal{P}}-\langle\hat{\mathcal{P}}\rangle_\epsilon$. According to Eq. \eqref{Bout}, we have
\begin{equation}
\begin{aligned}
\hat{\mathcal{X}}&=\frac{1}{\sqrt{\tau}}\int_0^\tau \frac{\beta+\beta^\dagger}{\sqrt{2}}+\frac{\hat{B}^{\textrm{in}}+\hat{B}^{\textrm{in}\dagger}}{\sqrt{2}}+\sqrt{\kappa}\hat{x}_m dt,\\
\hat{\mathcal{P}}&=\frac{1}{\sqrt{\tau}}\int_0^\tau i\frac{-\beta+\beta^\dagger}{\sqrt{2}}+i\frac{-\hat{B}^{\textrm{in}}+\hat{B}^{\textrm{in}\dagger}}{\sqrt{2}}+\sqrt{\kappa}\hat{p}_m dt.
\end{aligned}
\end{equation}

By the definition of $\mu_\epsilon$, it can be seen that
\begin{equation}
\begin{aligned}
\frac{d\mu_\epsilon}{d\epsilon}&=\sqrt{\kappa\tau}(\frac{d\langle\hat{x}_m\rangle_\epsilon}{d\epsilon},
\frac{d\langle\hat{p}_m\rangle_\epsilon}{d\epsilon})
\end{aligned}
\end{equation}
scales as $\beta$ owing to the $\beta$ dependance of $\frac{d\langle\hat{x}_m\rangle_\epsilon}{d\epsilon} $ and $\frac{d\langle\hat{p}_m\rangle_\epsilon}{d\epsilon}$. However, noting that
\begin{equation}
\begin{aligned}
\delta\hat{\mathcal{X}}&=\frac{1}{\sqrt{\tau}}\int_0^\tau\frac{\hat{B}^{\textrm{in}}+\hat{B}^{\textrm{in}\dagger}}{\sqrt{2}}+\sqrt{\kappa}(\hat{x}_m-\langle\hat{x}_m\rangle_\epsilon)dt,\\
\delta\hat{\mathcal{P}}&=\frac{1}{\sqrt{\tau}}\int_0^\tau i\frac{-\hat{B}^{\textrm{in}}+\hat{B}^{\textrm{in}\dagger}}{\sqrt{2}}+\sqrt{\kappa}(\hat{p}_m-\langle\hat{p}_m\rangle_\epsilon)dt,
\end{aligned}
\end{equation}
which are only determined by the noises. Thus, $V_{\epsilon}$  and  the second part of the QFI in Eq. (\ref{QFIQFI}) are independent of $\beta$.  Therefore,  in the large drive limit ($|\beta|\gg1$), QFI is determined by the first part of  Eq. \eqref{QFIQFI}.

Since $\epsilon$ is infinitesimal, the dominant term $\text{QFI}_{\text{D}}$ of QFI is
\begin{equation}\label{QFID}
\begin{aligned}
\text{QFI}_{\text{D}}=\frac{d\mu_\epsilon}{d\epsilon}V^{-1}_\epsilon\frac{d\mu_\epsilon^\top}{d\epsilon}\Bigg|_{\epsilon=0}.
\end{aligned}
\end{equation}
According to Eq. \eqref{signal} and Eq. \eqref{noise}, we have
\begin{equation}
\begin{aligned}
\mathcal{S}_\tau(N,\epsilon)=\kappa\tau\{\textrm{cos}\phi(\langle\hat{x}_m\rangle_\epsilon-\langle\hat{x}_m\rangle_0)+\textrm{sin}\phi(\langle\hat{p}_m\rangle_\epsilon-\langle\hat{p}_m\rangle_0)\}^2,
\end{aligned}
\end{equation}
\begin{equation}
\begin{aligned}
\mathcal{N}_\tau(N,\epsilon)=\langle(\textrm{cos}\phi\delta\hat{\mathcal{X}}+\textrm{sin}\phi\delta\hat{\mathcal{P}})^2\rangle_\epsilon.
\end{aligned}
\end{equation}
Since $\epsilon$ is infinitesimal, the dominant terms of $\mathcal{S}_\tau(N,\epsilon)$ and $\mathcal{N}_\tau(N,\epsilon)$
are
\begin{equation}\label{SNRD}
\begin{aligned}
\mathcal{S}_\tau(N,\epsilon)_\textrm{D}&=\kappa\tau\epsilon^2\{\textrm{cos}\phi\frac{d\langle\hat{x}_m\rangle_\epsilon}{d\epsilon}+\textrm{sin}\phi\frac{d\langle\hat{p}_m\rangle_\epsilon}{d\epsilon}\}^2=\epsilon^2\{\frac{d\mu_\epsilon}{d\epsilon}\begin{pmatrix}
\textrm{cos}\phi\\
\textrm{sin}\phi
\end{pmatrix}\}^2,\\
\mathcal{N}_\tau(N,\epsilon)_\textrm{D}&=\mathcal{N}_\tau(N,0)=(\textrm{cos}\phi,\textrm{sin}\phi)V_0\begin{pmatrix}
\textrm{cos}\phi\\
\textrm{sin}\phi
\end{pmatrix}.
\end{aligned}
\end{equation}
From Eq. (\ref{QFID}) and Eq. (\ref{SNRD}), it can be verified that  we have the following  relationship
\begin{equation}
\textrm{QFI}_{\text{D}}=\frac{1}{\epsilon^2}\frac{\mathcal{S}_\tau(N,\epsilon)_\textrm{D}}{\mathcal{N}_\tau(N,\epsilon)_\textrm{D}},
\end{equation}
by choosing appropriate measurement angle $\phi$, which depends on the phase and the position of the drive.
Interested readers can refer to \cite{Lau2018} for detailed derivations.

\section{Derivations of the SNR per photon with $\hat{V}_{\textsf{NHSE}}$}

According to the non-Hermitian Hamiltonian Eq.  \eqref{fullHamiltonian} and $\hat{V}_{\textsf{NHSE}}=e^{i\varphi}\hat{a}^\dagger _1\hat{a}_N+e^{-i\varphi}\hat{a}_1\hat{a}^\dagger_N$, the dynamics of the system can be described by the Heisenberg-Langevin equations as
\begin{equation}
\begin{aligned}
\dot{\hat{a}}_n=&w \hat{a}_{n-1}+\Delta\hat{a}^\dagger_{n+1}+\Delta\hat{a}^\dagger_{n-1}-w\hat{a}_{n+1}-\frac{\kappa}{2}\hat{a}_m\delta_{nm}\\ &-i\epsilon(e^{i\varphi}\hat{a}_N\delta_{n1}+e^{-i\varphi}\hat{a}_1\delta_{nN}) -\sqrt{\kappa}(|\beta|e^{i\theta}+\hat{B}^{\textsf{in}})\delta_{nm}.
\end{aligned}
\end{equation}
The corresponding  equations in terms of ${\hat{x}}_n$ and ${\hat{p}}_n$ are
\begin{equation}\label{XnPnSE}
\begin{aligned}
\dot{\hat{x}}_n=&-(w-\Delta)\hat{x}_{n+1}+(w+\Delta)\hat{x}_{n-1}+\epsilon \textrm{cos}\varphi(\hat{p}_N\delta_{n1}+\hat{p}_1\delta_{nN})-\epsilon \textrm{sin}\varphi(\hat{x}_1\delta_{nN}-\hat{x}_N\delta_{n1})\\ &-\frac{\kappa}{2}\hat{x}_m\delta_{nm}-\sqrt{2\kappa}|\beta|\textrm{cos}\theta\delta_{nm}-\sqrt{\kappa}\hat{X}^{\textrm{in}}\delta_{nm},\\
\dot{\hat{p}}_n=&(w-\Delta)\hat{p}_{n-1}-(w+\Delta)\hat{p}_{n+1}-\epsilon \textrm{cos}\varphi(\hat{x}_N\delta_{n1}+\hat{x}_1\delta_{nN})-\epsilon \textrm{sin}\varphi(\hat{p}_1\delta_{nN}-\hat{p}_N\delta_{n1})\\
&-\frac{\kappa}{2}\hat{p}_m\delta_{nm}-\sqrt{2\kappa}|\beta|\textrm{sin}\theta\delta_{nm}-\sqrt{\kappa}\hat{P}^{\textrm{in}}\delta_{nm},
\end{aligned}
\end{equation}
where $\hat{X}^{\textsf{in}}$ and $\hat{P}^{\textsf{in}}$ are defined via $\hat{B}^{\textsf{in}}=\frac{\hat{X}^{\textsf{in}}+i\hat{P}^{\textsf{in}}}{\sqrt{2}}$. Their average values are zero, and their second moments satisfy
\begin{equation}
\langle \hat{X}^{\textsf{in}}(t)\hat{X}^{\textsf{in}}(t')\rangle=(\bar{n}_{\textsf{th}}+\frac{1}{2})\delta(t-t'),
\end{equation}
\begin{equation}
\langle \hat{P}^{\textsf{in}}(t)\hat{P}^{\textsf{in}}(t')\rangle=(\bar{n}_{\textsf{th}}+\frac{1}{2})\delta(t-t'),
\end{equation}
\begin{equation}
\frac{1}{2}\langle \{ \hat{X}^{\textsf{in}}(t),\hat{P}^{\textsf{in}}(t')\}\rangle=0,
\end{equation}
where $\bar{n}_{\textsf{th}}$  is the number of thermal quanta in the input field. From Eq.~\eqref{JA}, Eq.  \eqref{XnPnSE} can be simplified as
\begin{equation}
\begin{aligned}
\dot{\hat{x}}_n=&-Je^{-A}\hat{x}_{n+1}+Je^{A}\hat{x}_{n-1}+\epsilon \textrm{cos}\varphi(\hat{p}_N\delta_{n1}+\hat{p}_1\delta_{nN})-\epsilon \textrm{sin}\varphi(\hat{x}_1\delta_{nN}-\hat{x}_N\delta_{n1})\\ &-\frac{\kappa}{2}\hat{x}_m\delta_{nm}-\sqrt{2\kappa}|\beta|\textrm{cos}\theta\delta_{nm}-\sqrt{\kappa}\hat{X}^{\textrm{in}}\delta_{nm},\\
\dot{\hat{p}}_n=&Je^{-A}\hat{p}_{n-1}-Je^A\hat{p}_{n+1}-\epsilon \textrm{cos}\varphi(\hat{x}_N\delta_{n1}+\hat{x}_1\delta_{nN})-\epsilon \textrm{sin}\varphi(\hat{p}_1\delta_{nN}-\hat{p}_N\delta_{n1})\\
&-\frac{\kappa}{2}\hat{p}_m\delta_{nm}-\sqrt{2\kappa}|\beta|\textrm{sin}\theta\delta_{nm}-\sqrt{\kappa}\hat{P}^{\textrm{in}}\delta_{nm}.
\end{aligned}
\end{equation}
The analysis can be further simplified by defining new canonically conjugate quadrature operators $\hat{\tilde{x}}_n$ and $\hat{\tilde{p}}_n$ by
\begin{equation}\label{squzeeing}
\begin{aligned}
\hat{\tilde{x}}_n&=e^{-A(n-1)}\hat{x}_n,\\
\hat{\tilde{p}}_n&=e^{A(n-1)}\hat{p}_n.
\end{aligned}
\end{equation}
The Heisenberg-Langevin equations of motion in terms of $\hat{\tilde{x}}_n$ and $\hat{\tilde{p}}_n$ are
\begin{equation}\label{XnPnSEsqueezing}
\begin{aligned}
\dot{\hat{\tilde{x}}}_n=&J\hat{\tilde{x}}_{n-1}-J\hat{\tilde{x}}_{n+1}+\epsilon \textrm{cos}\varphi(\hat{\tilde{p}}_Ne^{-A(N-1)}\delta_{n1}+\hat{\tilde{p}}_1e^{-A(N-1)}\delta_{nN})\\
&-\epsilon\textrm{sin}\varphi(\hat{\tilde{x}}_1e^{-A(N-1)}\delta_{nN}-\hat{\tilde{x}}_Ne^{A(N-1)}\delta_{n1})\\ &-\frac{\kappa}{2}\hat{\tilde{x}}_m\delta_{nm}-\sqrt{2\kappa}|\beta|\textrm{cos}\theta e^{-A(m-1)}\delta_{nm}-\sqrt{\kappa}\hat{X}^{\textrm{in}}e^{-A(m-1)}\delta_{nm},\\
\dot{\hat{\tilde{p}}}_n=&J\hat{\tilde{p}}_{n-1}-J\hat{\tilde{p}}_{n+1}-\epsilon \textrm{cos}\varphi(\hat{\tilde{x}}_Ne^{A(N-1)}\delta_{n1}+\hat{\tilde{x}}_1e^{A(N-1)}\delta_{nN})\\
&-\epsilon\textrm{sin}\varphi(\hat{\tilde{p}}_1e^{A(N-1)}\delta_{nN}-\hat{\tilde{p}}_Ne^{-A(N-1)}\delta_{n1})\\
&-\frac{\kappa}{2}\hat{\tilde{p}}_m\delta_{nm}-\sqrt{2\kappa}|\beta|\textrm{sin}\theta e^{A(m-1)}\delta_{nm}-\sqrt{\kappa}\hat{P}^{\textrm{in}}e^{A(m-1)}\delta_{nm}.
\end{aligned}
\end{equation}
Since Eq.~\eqref{XnPnSEsqueezing} is stable,  it is convenient to transfer into the frequency domain to solve Eq.~\eqref{XnPnSEsqueezing}. Define
\begin{equation}
\begin{aligned}
\tilde{\chi}[\omega,\epsilon]\triangleq (\omega \mathbb{I}-\mathbb{\tilde{H}}[\epsilon])^{-1},
\end{aligned}
\end{equation}
where $\mathbb{I}$ is the $2N \times 2N$ identity matrix. If the perturbation Hamiltonian is $\hat{V}_{\textsf{NHSE}}$,
\begin{equation}
\mathbb{\tilde{H}}[\epsilon]=\mathbb{\tilde{H}}_1(\kappa)+\mathbb{\tilde{H}}_{\textsf{NHSE}}(\epsilon),
\end{equation}
with
\begin{equation}\label{H1HSE}
\begin{aligned}
\mathbb{\tilde{H}}_1(\kappa)=&J\sum^{N-1}_{n=1} \left(|n+1\rangle\langle n|-|n \rangle\langle n+1| \right)+J\sum^{2N-1}_{n=N+1} \left(|n+1\rangle\langle n|-|n \rangle\langle n+1| \right)\\
&-\frac{\kappa}{2}|m\rangle\langle m|-\frac{\kappa}{2}|N+m\rangle\langle N+m|,\\
\mathbb{\tilde{H}}_{\textsf{NHSE}}(\epsilon)=&\epsilon\textrm{sin}\varphi e^{A(N-1)}|1\rangle\langle N|+\epsilon\textrm{cos}\varphi e^{-A(N-1)}|1\rangle\langle 2N|\\
&-\epsilon\textrm{sin}\varphi e^{-A(N-1)}|N\rangle\langle 1|+\epsilon\textrm{cos}\varphi e^{-A(N-1)}|N\rangle\langle N+1|\\
&-\epsilon\textrm{cos}\varphi e^{A(N-1)}|N+1\rangle\langle N|+\epsilon\textrm{sin}\varphi e^{-A(N-1)}|N+1\rangle\langle 2N|\\ &-\epsilon\textrm{cos}\varphi e^{A(N-1)}|2N\rangle\langle 1|-\epsilon\textrm{sin}\varphi e^{A(N-1)}|2N\rangle\langle N+1|,
\end{aligned}
\end{equation}
where $|n\rangle$ denotes a position eigenket. Moreover, define the zero-frequency transfer matrix as
\begin{equation}
\begin{aligned}
\tilde{\chi}[0,\epsilon]\triangleq-\mathbb{\tilde{H}}^{-1}(\epsilon).
\end{aligned}
\end{equation}
After sufficiently large time $t$, the $\hat{\tilde{x}}_n$ and $\hat{\tilde{p}}_n$ can be described as
\begin{equation}\label{xnpnsolution}
\begin{aligned}
\hat{\tilde{x}}_n=&\mathbb{\tilde{H}}[\epsilon]^{-1}_{n,m}(\sqrt{2\kappa}|\beta|\textrm{cos}\theta+\sqrt{\kappa}\hat{X}^{\textsf{in}})e^{-A(m-1)} +\mathbb{\tilde{H}}[\epsilon]^{-1}_{n,N+m}(\sqrt{2\kappa}|\beta|\textrm{sin}\theta+\sqrt{\kappa}\hat{P}^{\textsf{in}})e^{A(m-1)},\\
\hat{\tilde{p}}_n=&\mathbb{\tilde{H}}[\epsilon]^{-1}_{N+n,m}(\sqrt{2\kappa}|\beta|\textrm{cos}\theta+\sqrt{\kappa}\hat{X}^{\textsf{in}})e^{-A(m-1)} +\mathbb{\tilde{H}}[\epsilon]^{-1}_{N+n,N+m}(\sqrt{2\kappa}|\beta|\textrm{sin}\theta+\sqrt{\kappa}\hat{P}^{\textsf{in}})e^{A(m-1)}.
\end{aligned}
\end{equation}
It is straightforward to obtain $\hat{x}_n$ and $\hat{p}_n$ by an inverse squeezing transformation.

According to the definition of the signal power in Eq.~\eqref{signala}, we have
\begin{equation}\label{signal1}
\begin{aligned}
\mathcal{S}_\tau(N,\epsilon)=2\kappa\tau\kappa|\beta|^2\cdot\Big{|}\textrm{Re}\big{\{} e^{-i\phi}\big[&(h^{-1}_{1,m}\epsilon\textrm{sin}\varphi h^{-1}_{m,N}e^{-A(N-1)}-h^{-1}_{m,1}\epsilon\textrm{sin}\varphi h^{-1}_{N,m}e^{A(N-1)})\textrm{cos}\theta\\
&-(h^{-1}_{m,1}\epsilon\textrm{cos}\varphi h^{-1}_{N,m}+h^{-1}_{1,m}\epsilon\textrm{cos}\varphi h^{-1}_{m,N})\textrm{sin}\theta e^{A(2m-N-1)}\\
&+i(h^{-1}_{m,1}\epsilon\textrm{cos}\varphi h^{-1}_{N,m}+h^{-1}_{1,m}\epsilon\textrm{cos}\varphi h^{-1}_{m,N})\textrm{cos}\theta e^{-A(2m-N-1)}\\
&-i(h^{-1}_{m,1}\epsilon\textrm{sin}\varphi h^{-1}_{N,m}e^{-A(N-1)}+h^{-1}_{1,m}\epsilon\textrm{sin}\varphi h^{-1}_{m,N}e^{A(N-1)})\textrm{sin}\theta
\big]\big{\}}\Big{|}^2,
\end{aligned}
\end{equation}
where
\begin{equation}\label{h}
\begin{aligned}
h=J\sum^{N-1}_{n=1} \left(|n+1\rangle\langle n|-|n \rangle\langle n+1| \right)-\frac{\kappa}{2}|m\rangle\langle m|,
\end{aligned}
\end{equation}
which is the first $N\times N$ principal diagonal block of $\mathbb{\tilde{H}}_1(\kappa)$ (See Appendix E).

For the noise power, from Eqs. (\ref{QFI})-(\ref{noise}), computing the QFI only requires the zeroth order  in $\epsilon$ when $\epsilon$ is infinitesimal. Thus, we only need to consider the system without perturbation. If $\epsilon=0$, then the two Hatano-Nelson chains are completely decoupled. Therefore, the output noise is only the noise introduced by the input field. This can also be verified by calculation. According to the definition of the noise power in Eq.~\eqref{noise} and Eq.~\eqref{M(N)}, we have
\begin{equation}
\begin{aligned}
&\hat{\mathcal{M}}_\tau(N)\Big|_{\epsilon=0}-\langle\hat{\mathcal{M}}_\tau(N)\rangle\Big|_{\epsilon=0}\\ =&\frac{1}{\sqrt{2\tau}}\int^\tau_0e^{-i\phi}\big{(}\hat{B}^{\textsf{in}}+\sqrt{\kappa}(\hat{a}_m-\langle\hat{a}_m\rangle)
\big|_{\epsilon=0}\big{)}+e^{i\phi}\big{(}\hat{B}^{\textsf{in}\dagger}+\sqrt{\kappa}(\hat{a}^\dagger_m
-\langle\hat{a}^\dagger_m\rangle)\big|_{\epsilon=0}\big{)}dt\\ =&\frac{1}{\sqrt{2\tau}}\int^\tau_0e^{-i\phi}(1+\kappa h^{-1}_{m,m})\hat{B}^{\textsf{in}}+e^{i\phi}(1+\kappa h^{-1}_{m,m})\hat{B}^{\textsf{in}\dagger}dt.
\end{aligned}
\end{equation}
Hence, the noise power is
\begin{equation}\label{shotnoise}
\begin{aligned}
\mathcal{N}_\tau(N,0)=&(1+\kappa h^{-1}_{m,m})^2 \frac{2\bar{n}_{\textsf{th}}+1}{2}\\=&\bar{n}_{\textsf{th}}+\frac{1}{2},
\end{aligned}
\end{equation}
with $\bar{n}_{\textsf{th}}$ representing the number of thermal quanta in the input field and $h^{-1}_{m,m}=-\frac{2}{\kappa}$.

Following a similar reason to the noise power, we only concern the zeroth order term of the total average number of photons $\bar{n}_{\textsf{tot}}$  with respect to $\epsilon$. According to the definition Eq.~\eqref{totalphoton} of the $\bar{n}_{\textsf{tot}}$, we have
\begin{equation}\label{phton1}
\begin{aligned}
\bar{n}_{\textsf{tot}}&=\kappa|\beta|^2 \sum^N_{n=1}h^2_{n,m}(\textrm{cos}^2\theta e^{2A(n-m)}+\textrm{sin}^2\theta e^{-2A(n-m)})\\
&=\bar{n}_{\textsf{tot,D}}\cdot Z(A),
\end{aligned}
\end{equation}
where $\bar{n}_{\textsf{tot,D}}=\kappa|\beta|^2(h^2_{N,m}\textrm{cos}^2\theta e^{2A(N-m)}+h^2_{1,m}\textrm{sin}^2\theta e^{2A(m-1)})$ and $Z(A)=1+O(e^{-2A})$.

%So the SNR per photon can be given by \eqref{signal1}, \eqref{shotnoise} and \eqref{phton1}
%\begin{equation}
%\begin{aligned}
%\overline{\textrm{SNR}}=
%\end{aligned}
%\end{equation}

\section{Derivations of the SNR per photon with $\hat{V}_{N}$}
In this Appendix, we calculate the SNR with the perturbation Hamiltonian  $\hat{V}_{N}=\hat{a}^\dagger_{N}\hat{a}_{N}$. According to the non-Hermitian Hamiltonian Eq.  \eqref{fullHamiltonian}, the dynamics can be described by the Heisenberg-Langevin equations
\begin{equation}
\begin{aligned}
\dot{\hat{a}}_n=&w\hat{a}_{n-1}+\Delta\hat{a}^\dagger_{n+1}+\Delta\hat{a}^\dagger_{n-1}-w\hat{a}_{n+1}-\frac{\kappa}{2}\hat{a}_m\delta_{nm}\\ &-i\epsilon\hat{a}_N\delta_{nN} -\sqrt{\kappa}(|\beta|e^{i\theta}+\hat{B}^{\textsf{in}})\delta_{nm}.
\end{aligned}
\end{equation}
The corresponding  equations in terms of  ${\hat{x}}_n$ and ${\hat{p}}_n$ are
\begin{equation}
\begin{aligned}
\dot{\hat{x}}_n=&Je^A\hat{x}_{n-1}-Je^{-A}\hat{x}_{n+1}+\epsilon\hat{p}_N\delta_{nN}-\frac{\kappa}{2}\hat{x}_m\delta_{nm}-\sqrt{2\kappa}|\beta|\textrm{cos}\theta\delta_{nm}-\sqrt{\kappa}\hat{X}^{\textrm{in}}\delta_{nm},\\
\dot{\hat{p}}_n=&Je^{-A}\hat{p}_{n-1}-Je^{A}\hat{p}_{n+1}-\epsilon\hat{x}_N\delta_{nN}-\frac{\kappa}{2}\hat{p}_m\delta_{nm}-\sqrt{2\kappa}|\beta|\textrm{sin}\theta\delta_{nm}-\sqrt{\kappa}\hat{P}^{\textrm{in}}\delta_{nm}.
\end{aligned}
\end{equation}
After a similar squeezing transformation as Eq. \eqref{squzeeing}, we obtain the Heisenberg-Langevin equations
\begin{equation}
\begin{aligned}
\dot{\hat{\tilde{x}}}_n=&J\hat{\tilde{x}}_{n-1}-J\hat{\tilde{x}}_{n+1}+\epsilon\hat{\tilde{p}}_Ne^{-2A(N-1)}\delta_{nN}\\ &-\frac{\kappa}{2}\hat{\tilde{x}}_m\delta_{nm}-\sqrt{2\kappa}|\beta|\textrm{cos}\theta e^{-A(m-1)}\delta_{nm}-\sqrt{\kappa}\hat{X}^{\textrm{in}}e^{-A(m-1)}\delta_{nm},\\
\dot{\hat{\tilde{p}}}_n=&J\hat{\tilde{p}}_{n-1}-J\hat{\tilde{p}}_{n+1}-\epsilon\hat{\tilde{x}}_Ne^{2A(N-1)}\delta_{nN}\\
&-\frac{\kappa}{2}\hat{\tilde{p}}_m\delta_{nm}-\sqrt{2\kappa}|\beta|\textrm{sin}\theta e^{A(m-1)}\delta_{nm}-\sqrt{\kappa}\hat{P}^{\textrm{in}}e^{A(m-1)}\delta_{nm}.
\end{aligned}
\end{equation}

For sufficiently long time, $\hat{\tilde{x}}_n$ and $\hat{\tilde{p}}_n$  can be described as
\begin{equation}
\begin{aligned}
\hat{\tilde{x}}_n=&\mathbb{\tilde{H}}[\epsilon]^{-1}_{n,m}(\sqrt{2\kappa}|\beta|\textrm{cos}\theta+\sqrt{\kappa}\hat{X}^{\textsf{in}})e^{-A(m-1)} +\mathbb{\tilde{H}}[\epsilon]^{-1}_{n,N+m}(\sqrt{2\kappa}|\beta|\textrm{sin}\theta+\sqrt{\kappa}\hat{P}^{\textsf{in}})e^{A(m-1)},\\
\hat{\tilde{p}}_n=&\mathbb{\tilde{H}}[\epsilon]^{-1}_{N+n,m}(\sqrt{2\kappa}|\beta|\textrm{cos}\theta+\sqrt{\kappa}\hat{X}^{\textsf{in}})e^{-A(m-1)} +\mathbb{\tilde{H}}[\epsilon]^{-1}_{N+n,N+m}(\sqrt{2\kappa}|\beta|\textrm{sin}\theta+\sqrt{\kappa}\hat{P}^{\textsf{in}})e^{A(m-1)},
\end{aligned}
\end{equation}
where $\mathbb{\tilde{H}}[\epsilon]=\mathbb{\tilde{H}}_1(\kappa)+\mathbb{\tilde{H}}_N(\epsilon)$, and
\begin{equation}
\begin{aligned}
\mathbb{\tilde{H}}_N(\epsilon)=\epsilon e^{-2A(N-1)}|N\rangle\langle2N|-\epsilon e^{2A(N-1)}|2N\rangle\langle N|,
\end{aligned}
\end{equation}
is different from that in Eq.~\eqref{H1HSE} owing to with different perturbation Hamiltonians.

According to the definition of the signal power in Eq.~\eqref{signala}, we have
\begin{equation}
\begin{aligned}
\mathcal{S}_\tau(N,\epsilon)=2\kappa\tau\kappa|\beta|^2\cdot|h_{m,N}h_{N,m}|^2|\textrm{Re}\{e^{-i\phi}(-\epsilon\textrm{sin}\theta e^{-2A(N-m)}+i\epsilon\textrm{cos}\theta e^{2A(N-m)})\}|^2,
\end{aligned}
\end{equation}
where $h$ is the same as in Eq.~\eqref{h}.

For the noise power and the total average number of photons, as we only concern the zeroth order term in $\epsilon$,
 they are the same as in Eq.~\eqref{shotnoise} and Eq.~\eqref{phton1}, respectively.

\section{Calculation of $\mathbb{\tilde{H}}[\epsilon]^{-1}$}
In this Appendix, we take $\mathbb{\tilde{H}}[\epsilon]=\mathbb{\tilde{H}}_1(\kappa)+\mathbb{\tilde{H}}_{\textsf{NHSE}}(\epsilon)$ as an example to demonstrate how to compute the elements of $\mathbb{\tilde{H}}[\epsilon]^{-1}$.

Since the drive is coupled with $\hat{a}_m$ through a waveguide, the terms that we really concern are $\mathbb{\tilde{H}}[\epsilon]^{-1}_{m,m}$, $\mathbb{\tilde{H}}[\epsilon]^{-1}_{N+m,m}$, $\mathbb{\tilde{H}}[\epsilon]^{-1}_{m,N+m}$ and $\mathbb{\tilde{H}}[\epsilon]^{-1}_{N+m,N+m}$.

Using Dyson's equation and keeping it up to the first order in $\epsilon$, we have
\begin{equation}\label{HH1HSE}
\begin{aligned}
{[\mathbb{\tilde{H}}(\epsilon)]}^{-1}={[{\mathbb{\tilde{H}}_1(\kappa) +\mathbb{\tilde{H}}_{\textsf{NHSE}}(\epsilon)}]}^{-1} = [\mathbb{\tilde{H}}_1(\kappa)]^{-1}-[\mathbb{\tilde{H}}_1(\kappa)]^{-1}\mathbb{\tilde{H}}_{\textsf{NHSE}}
(\epsilon)[\mathbb{\tilde{H}}_1(\kappa)]^{-1}.
\end{aligned}
\end{equation}
Note that $\mathbb{\tilde{H}}_1(\kappa)$ is a block diagonal matrix consisting of two identical matrices $h$. Hence, the off-diagonal block elements of $[\mathbb{\tilde{H}}_1(\kappa)]^{-1}$ are 0. Multiplying Eq.~\eqref{HH1HSE} from left by $\langle m|$ and from right by $|m\rangle$ yields
\begin{equation}
\begin{aligned}
\mathbb{\tilde{H}}[\epsilon]^{-1}_{m,m}&= \langle m|[\mathbb{\tilde{H}}_1(\kappa)]^{-1}|m\rangle-\langle m|[\mathbb{\tilde{H}}_1(\kappa)]^{-1}\mathbb{\tilde{H}}_{\textsf{NHSE}}(\epsilon)[\mathbb{\tilde{H}}_1(\kappa)]^{-1}|m\rangle\\
&=h^{-1}_{m,m}-h^{-1}_{m,1}h^{-1}_{N,m}\epsilon\textrm{sin}\varphi e^{A(N-1)}+ h^{-1}_{m,N}h^{-1}_{1,m}\epsilon\textrm{sin}\varphi e^{-A(N-1)}.
\end{aligned}
\end{equation}
Other elements can be computed in a similar way. Now we just need to calculate the elements of $h$. Each column of $h$ has a recursive relationship. Let us take the first column of $h$ as an example. The other columns can be computed similarly.

According to the definition of $h$, we have
\begin{equation}
\begin{aligned}
\mathbb{I}=\left(J\sum^{N-1}_{n=1} \left(|n+1\rangle\langle n|-|n \rangle\langle n+1| \right)-\frac{\kappa}{2}|m\rangle\langle m|\right)h^{-1},
\end{aligned}
\end{equation}
where $\mathbb{I}$ is the $N\times N$ identity matrix. Multiplying this equation from left by $\langle i|$ and right by $|1\rangle$ for $i=1, \cdots, N$, yields
\begin{equation}
\begin{aligned}
1=&-J\langle2|h^{-1}|1\rangle,\\
0=&(J\langle1|-J\langle3|)h^{-1}|1\rangle,\\
0=&(J\langle2|-J\langle4|)h^{-1}|1\rangle,\\
\vdots\\
0=&(J\langle m-2|-J\langle m|)h^{-1}|1\rangle,\\
0=&(J\langle m-1|-J\langle m+1|-\frac{\kappa}{2}\langle m|)h^{-1}|1\rangle,\\
0=&(J\langle m|-J\langle m+2|)h^{-1}|1\rangle,\\
\vdots\\
0=&(J\langle N-2|-J\langle N|)h^{-1}|1\rangle,\\
0=&J\langle N-1|h^{-1}|1\rangle.
\end{aligned}
\end{equation}
Simplifying the above recursive formula, we obtain
\begin{equation}
\begin{aligned}
h^{-1}_{1,1}&=h^{-1}_{3,1}=\cdots=h^{-1}_{m,1}=\cdots=h^{-1}_{N,1}=-\frac{2}{\kappa};\\
h^{-1}_{2,1}&=h^{-1}_{4,1}=\cdots=h^{-1}_{m-1,1}=-\frac{1}{J};\\
h^{-1}_{m+1,1}&=h^{-1}_{m+3,1}=\cdots=h^{-1}_{N-1,1}=0.
\end{aligned}
\end{equation}
The above method is cumbersome but effective.

\section{All the orders of $\mathbb{\tilde{H}}[\epsilon_0]^{-1}$ with respect to $\epsilon_0$}
If $\epsilon_0$ is not infinitesimal, we have to compute all the orders with respect to $\epsilon_0$. However, the method to compute the elements of $\mathbb{\tilde{H}}[\epsilon_0]^{-1}$ is similar to what we have done in Appendix E.

According to the definition of $\mathbb{\tilde{H}}[\epsilon_0]$, we have
\begin{equation}
\begin{aligned}
\mathbb{I}=&\Bigg(J\sum^{N-1}_{n=1} \left(|n+1\rangle\langle n|-|n \rangle\langle n+1| \right)+J\sum^{2N-1}_{n=N+1} \left(|n+1\rangle\langle n|-|n \rangle\langle n+1| \right)\\ &-\frac{\kappa}{2}|N\rangle\langle N|-\frac{\kappa}{2}|2N\rangle\langle 2N|\\ &+\epsilon_0\textrm{sin}\varphi e^{A(N-1)}|1\rangle\langle N|+\epsilon_0\textrm{cos}\varphi e^{-A(N-1)}|1\rangle\langle 2N|\\ &-\epsilon_0\textrm{sin}\varphi e^{-A(N-1)}|N\rangle\langle 1|+\epsilon_0\textrm{cos}\varphi e^{-A(N-1)}|N\rangle\langle N+1|\\ &-\epsilon_0\textrm{cos}\varphi e^{A(N-1)}|N+1\rangle\langle N|+\epsilon_0\textrm{sin}\varphi e^{-A(N-1)}|N+1\rangle\langle 2N|\\ &-\epsilon_0\textrm{cos}\varphi e^{A(N-1)}|2N\rangle\langle 1|-\epsilon_0\textrm{sin}\varphi e^{A(N-1)}|2N\rangle\langle N+1| \Bigg)\mathbb{\tilde{H}}[\epsilon_0]^{-1}.
\end{aligned}
\end{equation}
where $\mathbb{I}$ is the $2N\times 2N$ identity matrix. Multiplying this equation from left by $\langle i|$ and right by $|1\rangle$ for $i=1, \cdots, N$, yields
\begin{equation}
\begin{aligned}
1&=\left(-J\langle2|+\epsilon_0\textrm{sin}\varphi e^{A(N-1)}\langle N|+\epsilon_0\textrm{cos}\varphi e^{-A(N-1)}\langle 2N|\right)\mathbb{\tilde{H}}[\epsilon_0]^{-1}|1\rangle,\\
0&=(J\langle1|-J\langle3|)\mathbb{\tilde{H}}[\epsilon_0]^{-1}|1\rangle,\\
0&=(J\langle2|-J\langle4|)\mathbb{\tilde{H}}[\epsilon_0]^{-1}|1\rangle,\\
\vdots\\
0&=(J\langle N-2|-J\langle N|)\mathbb{\tilde{H}}[\epsilon_0]^{-1}|1\rangle,\\
0&=(J\langle N-1|-\frac{\kappa}{2}\langle N|-\epsilon_0\textrm{sin}\varphi e^{-A(N-1)}\langle1|+\epsilon_0\textrm{cos}\varphi e^{-A(N-1)}\langle N+1|)\mathbb{\tilde{H}}[\epsilon_0]^{-1}|1\rangle,\\
0&=(-J\langle N+2|-\epsilon_0\textrm{cos}\varphi e^{A(N-1)}\langle N|+\epsilon_0\textrm{sin}\varphi e^{-A(N-1)}\langle 2N|)\mathbb{\tilde{H}}[\epsilon_0]^{-1}|1\rangle,\\
\vdots\\
0&=(J\langle 2N-1|-\frac{\kappa}{2}\langle 2N|-\epsilon_0\textrm{cos}\varphi e^{A(N-1)}\langle 1|-\epsilon_0\textrm{sin}\varphi e^{A(N-1)}\langle N+1|)\mathbb{\tilde{H}}[\epsilon_0]^{-1}|1\rangle.
\end{aligned}
\end{equation}
Simplifying the above recursive formula, we obtain
\begin{equation}
\begin{aligned}
\mathbb{\tilde{H}}[\epsilon_0]^{-1}_{1,1}&=\mathbb{\tilde{H}}[\epsilon_0]^{-1}_{3,1}=\cdots=\mathbb{\tilde{H}}[\epsilon_0]^{-1}_{N,1}\\ &=\frac{2\kappa e^{A(N-1)}+4e^{A(N-1)}(-1+e^{2A(N-1)})\epsilon_0 \textrm{sin}\varphi}{-e^{2A(N-1)}\kappa^2-16e^{2A(N-1)}\epsilon_0^2\textrm{cos}^2\varphi+4(-1+e^{2A(N-1)})^2\epsilon_0^2\textrm{sin}^2\varphi};\\
\mathbb{\tilde{H}}[\epsilon_0]^{-1}_{2,1}&=\mathbb{\tilde{H}}[\epsilon_0]^{-1}_{4,1}=\cdots=\mathbb{\tilde{H}}[\epsilon_0]^{-1}_{N-1,1}\\ &=\frac{e^{2A(N-1)}\kappa^2+8e^{2A(N-1)}\epsilon_0^2\textrm{cos}^2\varphi+2e^{3A(N-1)}\kappa\epsilon_0\textrm{sin}\varphi+4(-1+e^{2A(N-1)})\epsilon_0^2 \textrm{sin}^2\varphi}{J(-e^{2A(N-1)}\kappa^2-16e^{2A(N-1)}\epsilon_0^2\textrm{cos}^2\varphi+4(-1+e^{2A(N-1)})^2\epsilon_0^2\textrm{sin}^2\varphi)};\\
\mathbb{\tilde{H}}[\epsilon_0]^{-1}_{N+2,1}&=\mathbb{\tilde{H}}[\epsilon_0]^{-1}_{N+4,1}=\cdots=\mathbb{\tilde{H}}[\epsilon_0]^{-1}_{2N-1,1}\\ &=-\frac{2e^{2A(N-1)}\epsilon_0\textrm{cos}\varphi(e^{A(N-1)}\kappa+2(1+e^{2A(N-1)})\epsilon_0\textrm{sin}\varphi)}{J(-e^{2A(N-1)}\kappa^2-16e^{2A(N-1)}\epsilon_0^2\textrm{cos}^2\varphi+4(-1+e^{2A(N-1)})^2\epsilon_0^2\textrm{sin}^2\varphi)};\\
\mathbb{\tilde{H}}[\epsilon_0]^{-1}_{N+1,1}&=\mathbb{\tilde{H}}[\epsilon_0]^{-1}_{N+3,1}=\cdots=\mathbb{\tilde{H}}[\epsilon_0]^{-1}_{2N,1}\\ &=-\frac{8e^{3A(N-1)}\epsilon_0\textrm{cos}\varphi}{-e^{2A(N-1)}\kappa^2-16e^{2A(N-1)}\epsilon_0^2\textrm{cos}^2\varphi+4(-1+e^{2A(N-1)})^2\epsilon_0^2\textrm{sin}^2\varphi}.
\end{aligned}
\end{equation}
The other columns can be computed in a similar way.

\end{widetext}


\begin{thebibliography}{0}%
\makeatletter
\providecommand \@ifxundefined [1]{%
 \@ifx{#1\undefined}
}%
\providecommand \@ifnum [1]{%
 \ifnum #1\expandafter \@firstoftwo
 \else \expandafter \@secondoftwo
 \fi
}%
\providecommand \@ifx [1]{%
 \ifx #1\expandafter \@firstoftwo
 \else \expandafter \@secondoftwo
 \fi
}%
\providecommand \natexlab [1]{#1}%
\providecommand \enquote  [1]{``#1''}%
\providecommand \bibnamefont  [1]{#1}%
\providecommand \bibfnamefont [1]{#1}%
\providecommand \citenamefont [1]{#1}%
\providecommand \href@noop [0]{\@secondoftwo}%
\providecommand \href [0]{\begingroup \@sanitize@url \@href}%
\providecommand \@href[1]{\@@startlink{#1}\@@href}%
\providecommand \@@href[1]{\endgroup#1\@@endlink}%
\providecommand \@sanitize@url [0]{\catcode `\\12\catcode `\$12\catcode
  `\&12\catcode `\#12\catcode `\^12\catcode `\_12\catcode `\%12\relax}%
\providecommand \@@startlink[1]{}%
\providecommand \@@endlink[0]{}%
\providecommand \url  [0]{\begingroup\@sanitize@url \@url }%
\providecommand \@url [1]{\endgroup\@href {#1}{\urlprefix }}%
\providecommand \urlprefix  [0]{URL }%
\providecommand \Eprint [0]{\href }%
\providecommand \doibase [0]{http://dx.doi.org/}%
\providecommand \selectlanguage [0]{\@gobble}%
\providecommand \bibinfo  [0]{\@secondoftwo}%
\providecommand \bibfield  [0]{\@secondoftwo}%
\providecommand \translation [1]{[#1]}%
\providecommand \BibitemOpen [0]{}%
\providecommand \bibitemStop [0]{}%
\providecommand \bibitemNoStop [0]{.\EOS\space}%
\providecommand \EOS [0]{\spacefactor3000\relax}%
\providecommand \BibitemShut  [1]{\csname bibitem#1\endcsname}%
\let\auto@bib@innerbib\@empty
%</preamble>
\end{thebibliography}%


\begin{thebibliography}{1}

%1-10

\bibitem{Monifi2016}
F. Monifi, J. Zhang, $\c{S}$. K. $\ddot{\textrm{O}}$zdemir, B. Peng, Y.-X. Liu, F. Bo, F. Nori and L. Yang, Optomechanically induced stochastic resonance and chaos transfer between optical fields, \href{https://www.nature.com/articles/nphoton.2016.73}{Nat. Photon. \textbf{10}, 399-405 (2016).}

\bibitem{ZhangJing2018}
J. Zhang, B. Peng, $\c{\textrm{S}}$. K. $\ddot{\textrm{O}}$zdemir, K. Pichler, D. O. Krimer, G. Zhao, F. Nori, Y.-X. Liu, S. Rotter and L. Yang, A phonon laser operating at an exceptional point, \href{https://www.nature.com/articles/s41566-018-0213-5/#citeas}{Nat. Photon. \textbf{12}, 479-484 (2018).}




\bibitem{Okuma2020}
N. Okuma, K. Kawabata, K. Shiozaki and M. Sato, Topological origin of non-Hermitian skin effects, \href{https://journals.aps.org/prl/abstract/10.1103/PhysRevLett.124.086801}{Phys. Rev. Lett. \textbf{124}, 086801 (2020).}

\bibitem{Kawabata2020}
K. Kawabata, N. Okuma and M. Sato, Non-bloch band theory of non-Hermitian Hamiltonians in the symplectic class, \href{https://journals.aps.org/prb/abstract/10.1103/PhysRevB.101.195147}{Phys. Rev. B \textbf{101}, 195147 (2020).}

\bibitem{Budich2020}
J. C. Budich and E. J. Bergholtz, Non-Hermitian topological sensors, \href{https://journals.aps.org/prl/abstract/10.1103/PhysRevLett.125.180403}{Phys. Rev. Lett. \textbf{125}, 180403 (2020).}

\bibitem{Koch2021}
F. Koch and J. C. Budich, Quantum non-Hermitian topological sensors, 	arXiv:2106.05297 (2021).

\bibitem{McDonald2020}
A. McDonald and A. A. Clerk, Exponentially-enhanced quantum sensing with non-Hermitian lattice dynamics, \href{https://www.nature.com/articles/s41467-020-19090-4}{ Nat. Commun. \textbf{11}, 5382 (2020).}

\bibitem{Rao2021}
J. Rao, Y. Zhao, Y. Gui, X. Fan, D. Xue and C.-M. Hu, Controlling microwaves in non-Hermitian metamaterials, \href{https://journals.aps.org/prapplied/abstract/10.1103/PhysRevApplied.15.L021003}{Phys. Rev. Applied \textbf{15}, L021003 (2021).}

\bibitem{Park2021}
S. Park, J. Lee and S. Kim, Robust wireless power transfer with minimal field exposure using parity-time symmetric microwave cavities, \href{https://journals.aps.org/prapplied/abstract/10.1103/PhysRevApplied.16.014022}{Phys. Rev. Applied \textbf{16}, 014022 (2021).}

\bibitem{Zhong2020}
F. Zhong, K. Ding, Y. Zhang, S. Zhu, C. Chan and H. Liu, Angle-resolved thermal emission spectroscopy characterization of non-Hermitian metacrystals, \href{https://journals.aps.org/prapplied/abstract/10.1103/PhysRevApplied.13.014071}{Phys. Rev. Applied \textbf{13}, 014071 (2020).}



\bibitem{Bao2021}
L. Bao, B. Qi, D. Dong and F. Nori, Fundamental limits for reciprocal and nonreciprocal non-Hermitian quantum sensing, \href{https://journals.aps.org/pra/abstract/10.1103/PhysRevA.103.042418}{ Phys. Rev. A \textbf{103}, 042418 (2021).}

\bibitem{Bensa2021}
J. Bensa and M. $\breve{\textrm{Z}}$nidari$\breve{\textrm{c}}$, Fastest local entanglement scrambler, multistage thermalization, and a non-Hermitian phantom, \href{https://journals.aps.org/prx/abstract/10.1103/PhysRevX.11.031019}{Phys. Rev. X \textbf{11}, 031019 (2021).}

\bibitem{G.-Q.Zhang2021}
G.-Q. Zhang, Z. Chen, D. Xu, N. Shammah, M. Liao, T.-F. Li, L. Tong, S.-Y. Zhu, F. Nori and J. Q. You, Exceptional point and cross-relaxation effect in a hybrid quantum system, \href{https://journals.aps.org/prxquantum/abstract/10.1103/PRXQuantum.2.020307}{PRX Quantum \textbf{2}, 020307 (2021).}

\bibitem{Howl2021}
R. Howl, V. Vedral, D. Naik, M. Christodoulou, C. Rovelli and A. Iyer, Non-Gaussianity as a signature of a quantum theory of gravity, \href{https://journals.aps.org/prxquantum/abstract/10.1103/PRXQuantum.2.010325}{PRX Quantum \textbf{2}, 010325 (2021).}

\bibitem{Roberts2021}
D. Roberts, A. Lingenfelter and A. A. Clerk, Hidden time-reversal symmetry, quantum detailed balance and exact solutions of driven-dissipative quantum systems, \href{https://journals.aps.org/prxquantum/abstract/10.1103/PRXQuantum.2.020336}{PRX Quantum \textbf{2}, 020336 (2021).}

\bibitem{Xiao2021}
L. Xiao, D. Qu, K. Wang, H.-W. Li, J.-Y. Dai, B. D$\acute{\textrm{o}}$ra, M. Heyl, R. Moessner, W. Yi and P. Xue, Non-Hermitian Kibble-Zurek Mechanism with tunable complexity in single-photon interferometry, \href{https://journals.aps.org/prxquantum/abstract/10.1103/PRXQuantum.2.020313}{PRX Quantum \textbf{2}, 020313 (2021).}


%11-20


\bibitem{Liu2019}
T. Liu, Y. R. Zhang, Q. Ai, Z. Gong, K. Kawabata, M. Ueda and F. Nori,
Second-order topological phases in non-Hermitian systems,
\href{https://journals.aps.org/prl/abstract/10.1103/PhysRevLett.122.076801}{ Phys. Rev. Lett. \textbf{122}, 076801 (2019).}

\bibitem{Bliokh2019}
K. Bliokh, D. Leykam, M. Lein and F. Nori, Topological non-Hermitian origin of surface Maxwell waves,  \href{https://www.nature.com/articles/s41467-019-08397-6#citeas}{  Nat. Commun. \textbf{10}, 580 (2019).}

\bibitem{Pan2020}
L. Pan, X. Chen, Y. Chen and H. Zhai, Non-Hermitian linear response theory, \href{https://www.nature.com/articles/s41567-020-0889-6}{Nat. Phys. \textbf{16}, 767-771 (2020).}

\bibitem{Cao2020}
W. Cao, X. Lu, X. Meng, J. Sun, H. Shen and Y. Xiao, Reservoir-mediated quantum correlations in non-Hermitian optical system, \href{https://journals.aps.org/prl/abstract/10.1103/PhysRevLett.124.030401}{Phys. Rev. Lett. \textbf{124}, 030401 (2020).}

\bibitem{Scheibner2020}
C. Scheibner, W. T. M. Irvine and V. Vitelli, Non-Hermitian band topology and skin modes in active elastic media, \href{https://journals.aps.org/prl/abstract/10.1103/PhysRevLett.125.118001}
{Phys. Rev. Lett. \textbf{125}, 118001 (2020).}

\bibitem{Ashida2020}
Y. Ashida, Z. Gong and M. Ueda, Non-Hermitian physics, arXiv:2006.01837 (2020).

\bibitem{Sun2020}
L. Sun, X. He, C. You, C. Lv, B. Li, S. Lloyd, and X. Wang, Exponential enhancement of quantum metrology using continuous variables, arXiv:2004.01216 (2020).



\bibitem{Gardiner2000}
C. W. Gardiner and P. Zoller, \textit{Quantum noise: A handbook of markovian and non-markovian quantum stochastic methods with applications to quantum optics, 2nd ed}. (Springer-Verlag, Berlin, 2000).

\bibitem{Ozdemir2014}
B. Peng, $\c{S}$. K. $\ddot{\textrm{O}}$zdemir, F. Lei, F. Monifi, M. Gianfreda, G. L. Long, S. Fan, F. Nori, C. M. Bender and L. Yang,
Parity-time-symmetric whispering-gallery microcavities, \href{https://www.nature.com/articles/nphys2927}{
Nat. Phys. \textbf{10}, 394-398 (2014).}

%20-30



\bibitem{Zhangjing2015}
J. Zhang, B. Peng, $\c{\textrm{S}}$. K. $\ddot{\textrm{O}}$zdemir, Y.-X. Liu, H. Jing, X.-Y. L$\ddot{\textrm{u}}$, Y.-L. Liu, L. Yang and F. Nori, Giant nonlinearity via breaking parity-time symmetry: a route to low-threshold phonon diodes, \href{https://journals.aps.org/prb/abstract/10.1103/PhysRevB.92.115407}{Phys. Rev. B \textbf{92}, 115407 (2015).}

\bibitem{Feng2017}
L. Feng, R. El-Ganainy and L. Ge, Non-{H}ermitian photonics based on parity-time symmetry, \href{https://www.nature.com/articles/s41566-017-0031-1}{ Nat. Photon. \textbf{11}, 752-762 (2017).}

\bibitem{Sounas2017}
D. L. Sounas and A. Al\`{u}, Non-reciprocal photonics based on time modulation, \href{https://www.nature.com/articles/s41566-017-0051-x}{ Nat. Photon. \textbf{11}, 774-783 (2017).}

\bibitem{Liuyl2017}
Y.-L. Liu, R. Wu, J. Zhang, $\c{\textrm{S}}$. K. $\ddot{\textrm{O}}$zdemir, L. Yang, F. Nori and Y.-X. Liu, Controllable optical response by modifying the gain and loss of a mechanical resonator and cavity mode in an optomechanical system, \href{https://journals.aps.org/pra/abstract/10.1103/PhysRevA.95.013843}{Phys. Rev. A \textbf{95}, 013843 (2017).}

\bibitem{El-Ganainy2018}
R. El-Ganainy, K. G. Makris, M. Khajavikhan, Z. H. Musslimani, S. Rotter and D. N. Christodoulides, Non-{H}ermitian physics and {PT} symmetry, \href{https://www.nature.com/articles/nphys4323}{ Nat. Phys. \textbf{14}, 11-19 (2018).}






\bibitem{Clerk2010}
A. A. Clerk, M. H. Devoret, S. M. Girvin, F. Marquardt and R. J. Schoelkopf, Introduction to quantum noise, measurement, and amplification, \href{https://journals.aps.org/rmp/abstract/10.1103/RevModPhys.82.1155}{ Rev. Mod. Phys. \textbf{82}, 1155-1208 (2010).}

\bibitem{Wiersig2014}
J. Wiersig, Enhancing the sensitivity of frequency and energy splitting detection by using exceptional points: application to microcavity sensors for
single-particle detection, \href{https://journals.aps.org/prl/abstract/10.1103/PhysRevLett.112.203901}{ Phys. Rev. Lett. \textbf{112}, 203901 (2014).}

\bibitem{Wiersig2016}
J. Wiersig, Sensors operating at exceptional points: general theory, \href{https://journals.aps.org/pra/abstract/10.1103/PhysRevA.93.033809}{ Phys. Rev. A \textbf{93}, 033809 (2016).}

\bibitem{Hodaei2017}
H. Hodaei, A. U. Hassan, S. Wittek, H. Garcia-Gracia, R. El-Ganainy, D. N. Christodoulides and M. Khajavikhan, Enhanced sensitivity at higher-order exceptional points, \href{https://www.nature.com/articles/nature23280#citeas} { Nature \textbf{548}, 187-191 (2017).}

\bibitem{Ren2017}
J. Ren, H. Hodaei, G. Harari, A. U. Hassan, W. Chow, M. Soltani, D. Christodoulides and M. Khajavikhan, Ultrasensitive micro-scale parity-time-symmetric ring laser
gyroscope, \href{https://www.osapublishing.org/ol/abstract.cfm?uri=ol-42-8-1556}{ Opt. Lett. \textbf{42}, 1556-1559 (2017).}

\bibitem{Chen2017}
W. Chen, $\c{S}$. K. $\ddot{\textrm{O}}$zdemir, G. Zhao, J. Wiersig and L. Yang, Exceptional points enhance sensing in an optical microcavity, \href{https://www.nature.com/articles/nature23281/#citeas}{ Nature \textbf{548}, 192-196 (2017).}



\bibitem{Langbein2018}
W. Langbein, No exceptional precision of exceptional-point sensors, \href{https://journals.aps.org/pra/abstract/10.1103/PhysRevA.98.023805}{ Phys. Rev. A \textbf{98}, 023805 (2018).}

\bibitem{Ozdemir2019}
$\c{S}$. K. $\ddot{\textrm{O}}$zdemir, S. Rotter, F. Nori and L. Yang, Parity-time symmetry and exceptional points in photonics, \href{https://www.nature.com/articles/s41563-019-0304-9}{ Nat. Mater.  \textbf{18}, 783-798 (2019).}

\bibitem{Zhang2019}
M. Zhang, W. Sweeney, C. W. Hsu, L. Yang, A. D. Stone and L. Jiang, Quantum noise theory of exceptional point amplifying sensors, \href{https://journals.aps.org/prl/abstract/10.1103/PhysRevLett.123.180501}{ Phys. Rev. Lett. \textbf{123}, 180501 (2019).}

\bibitem{Chen2019}
C. Chen, L. Jin and R.-B. Liu, Sensitivity of parameter estimation near the exceptional point of a non-Hermitian system, \href{https://iopscience.iop.org/article/10.1088/1367-2630/ab32ab/meta}{ New J. Phys. \textbf{21}, 083002 (2019).}

\bibitem{Lau2018}
H.-K. Lau and A. A. Clerk, Fundamental limits and non-reciprocal approaches in non-{H}ermitian quantum sensing, \href{https://www.nature.com/articles/s41467-018-06477-7#citeas}{  Nat. Commun. \textbf{9}, 4320 (2018).}


\bibitem{Demange2012}
G. Demange and E.-M. Graefe, Signatures of three coalescing eigenfunctions, \href{https://iopscience.iop.org/article/10.1088/1751-8113/45/2/025303/meta}{ J. Phys. A-Math. Theor. \textbf{45}, 025303 (2012).}

\bibitem{Jing2014}
H. Jing, $\c{\textrm{S}}$. K. $\ddot{\textrm{O}}$zdemir, X.-Y. L$\ddot{\textrm{u}}$, J. Zhang, L. Yang and F. Nori, PT-symmetric phonon laser, \href{https://journals.aps.org/prl/abstract/10.1103/PhysRevLett.113.053604}{Phys. Rev. Lett. \textbf{113}, 053604 (2014).}

\bibitem{Chen2016}
P.-Y. Chen and J. Jung, {PT}-symmetry and singularity-enhanced sensing based on photoexcited graphene metasurfaces, \href{https://journals.aps.org/prapplied/abstract/10.1103/PhysRevApplied.5.064018}{ Phys. Rev. Applied \textbf{5}, 064018 (2016).}

\bibitem{Liu2016}
Z.-P. Liu, J. Zhang, $\c{S}$. K. \"{O}zdemir, B. Peng, H. Jing, X.-Y. L\"{u}, C.-W. Li, L. Yang, F. Nori, and Y.-X. Liu, Metrology with {PT}-symmetric cavities: enhanced sensitivity
near the {PT}-phase transition, \href{https://journals.aps.org/prl/abstract/10.1103/PhysRevLett.117.110802}{ Phys, Rev. Lett. \textbf{117}, 110802 (2016).}

\bibitem{Metelmann2015}
A. Metelmann and A. A. Clerk, Non-reciprocal photon transmission and amplification via reservoir engineering, \href{https://journals.aps.org/prx/abstract/10.1103/PhysRevX.5.021025}{ Phys. Rev. X \textbf{5}, 021025 (2015).}


\bibitem{Huai2019}
S.-N. Huai, Y.-L. Liu, J. Zhang, L. Yan and Y.-X. Liu, Enhanced sideband responses in a PT-symmetric-like cavity magnomechanical system, \href{https://journals.aps.org/pra/abstract/10.1103/PhysRevA.99.043803}{Phys. Rev. A \textbf{99}, 043803 (2019).}



\bibitem{Chu2020}
Y. Chu, Y. Liu, H. Liu and J. Cai, Quantum sensing with a single-qubit pseudo-Hermitian system, \href{https://journals.aps.org/prl/abstract/10.1103/PhysRevLett.124.020501}{Phys. Rev. Lett. \textbf{124}, 020501 (2020).}




\bibitem{Dembowski2001}
C. Dembowski, H.-D. Gr$\ddot{\textrm{a}}$f, H. L. Harney, A. Heine, W. D. Heiss, H. Rehfeld and A. Richter, Experimental observation of the topological structure of exceptional points,  \href{https://journals.aps.org/prl/abstract/10.1103/PhysRevLett.86.787}{ Phys. Rev. Lett. \textbf{86}, 787-790 (2001).}

\bibitem{Heiss2004}
W. D. Heiss, Exceptional points of non-{H}ermitian operators, \href{https://iopscience.iop.org/article/10.1088/0305-4470/37/6/034}{ J. Phys. A-Math. Theor. \textbf{37}, 2455-2464 (2004).}

\bibitem{Seyranian2005}
A. P. Seyranian, O. N. Kirillov and A. A. Mailybaev, Coupling of eigenvalues of complex matrices at diabolic and exceptional points, \href{https://iopscience.iop.org/article/10.1088/0305-4470/38/8/009/meta}{ J. Phys. A-Math. Theor. \textbf{38}, 1723-1740 (2005).}


\bibitem{Liertzer2012}
M. Liertzer, L. Ge, A. Cerjan, A. D. Stone, H. E. T$\ddot{\textrm{u}}$reci and S. Rotter, Pump-induced exceptional points in lasers, \href{https://journals.aps.org/prl/abstract/10.1103/PhysRevLett.108.173901}{ Phys. Rev. Lett. \textbf{108}, 173901 (2012).}

\bibitem{Heiss2012}
W. D. Heiss, The physics of exceptional points, \href{https://iopscience.iop.org/article/10.1088/1751-8113/45/44/444016/meta}{ J. Phys. A-Math. Theor. \textbf{45}, 444016 (2012).}

%30-40

\bibitem{Rotter2014}
B. Peng, $\c{S}$. K. $\ddot{\textrm{O}}$zdemir, S. Rotter, H. Yilmaz, M. Liertzer, F. Monifi, C. M. Bender, F. Nori and L. Yang, Loss-induced suppression and revival of lasing, \href{https://science.sciencemag.org/content/346/6207/328}{
Science \textbf{346}, 328-332 (2014).}

\bibitem{Zhen2015}
B. Zhen, C. W. Hsu, Y. Igarashi, L. Lu, I. Kaminer, A. Pick, S.-L. Chua, J. D. Joannopoulos and M. Solja$\breve{\textrm{c}}$i$\acute{\textrm{c}}$, Spawning rings of exceptional points out of Dirac cones, \href{https://www.nature.com/articles/nature14889#auth-1}{ Nature \textbf{525}, 354-358 (2015).}

\bibitem{Xu2016}
H. Xu, D. Mason, L. Jiang and J. G. E. Harris, Topological energy transfer in an optomechanical system with exceptional points, \href{https://www.nature.com/articles/nature18604#auth-4}{ Nature \textbf{537}, 80-83 (2016).}

\bibitem{Sunada2017}
S. Sunada, Large Sagnac frequency splitting in a ring resonator operating at an exceptional point, \href{https://journals.aps.org/pra/abstract/10.1103/PhysRevA.96.033842}{ Phys. Rev. A \textbf{96}, 033842 (2017).}


\bibitem{Leykam2017}
D. Leykam, K. Y. Bliokh, C. Huang, Y. D. Chong and F. Nori, Edge modes, degeneracies, and topological numbers in non-Hermitian systems,  \href{https://journals.aps.org/prl/abstract/10.1103/PhysRevLett.118.040401}{ Phys. Rev. Lett. \textbf{118}, 040401 (2017).}

%40-50



\bibitem{Pinel2013}
O. Pinel, P. Jian, N. Treps, C. Fabre and D. Braun, Quantum parameter
estimation using general single-mode Gaussian states, \href{https://journals.aps.org/pra/abstract/10.1103/PhysRevA.88.040102}{ Phys. Rev. A \textbf{88}, 040102 (2013).}

\bibitem{Banchi2015}
L. Banchi, S. L. Braunstein and S. Pirandola, Quantum fidelity for arbitrary Gaussian states, \href{https://journals.aps.org/prl/abstract/10.1103/PhysRevLett.115.260501}{ Phys. Rev. Lett. \textbf{115}, 260501 (2015).}



\bibitem{OH2019}
C. Oh, C. Lee, C. Rockstuhl, H. Jeong, J. Kim, H. Nha and S. Lee, Optimal Gaussian measurements for phase estimation in single-mode Gaussian metrology, \href{https://www.nature.com/articles/s41534-019-0124-4#citeas}{npj Quantum Inf. \textbf{5}, 10 (2019).}



\end{thebibliography}
\end{document}